\documentclass[epj,amsfonts]{svjour}
\usepackage{graphics}
\usepackage{graphicx}% Include figure files
\usepackage{dcolumn}% Align table columns on decimal point
\usepackage{amssymb}
\usepackage[usenames]{color}
\usepackage{amsfonts}
\usepackage{amssymb}
\usepackage{amsmath}
\usepackage{mathrsfs}
\usepackage{mathrsfs}
\usepackage{amssymb}
\usepackage{comment}

\def\dbar{{\mathchar'26\mkern-12mu \mathrm{d}}}

\newcommand{\ar}{\arrowvert}
\newcommand{\ra}{\rangle}
\newcommand{\la}{\langle}
\newcommand{\da}{\dagger}

\newcommand{\be}{\begin{equation}}
\newcommand{\ee}{\end{equation}}
\newcommand{\ba}{\begin{eqnarray}}
\newcommand{\ea}{\end{eqnarray}}

\begin{document}
\title{Boost operators in Coulomb-gauge QCD:\\
the pion form factor and Fock expansions in $\phi$ radiative decays}
\author{
Mar\'{\i}a G\'omez Rocha$ ^1$, Felipe J. Llanes-Estrada$ ^2$, Dieter Sch\"utte$ ^3$ and Selym Villalba-Ch\'avez$ ^1$.
}                     % Do not remove
\institute{ 
$ ^1$ Institut f\"ur Physik, U. Graz,
Universit\"atsplatz 5, A-8010 Graz, Austria\\
$ ^2$ Departamento de F\'{\i}sica Te\'orica I,  Universidad
Complutense, 28040 Madrid, Spain \\
$ ^3$ Emeritus from U. Bonn, Institut fur Theoretische Kernphysik,
Nussallee 14-16, D-53115 Bonn, Germany
}
\date{Received: date / Revised version: date}

\abstract{
In this article we rederive the Boost operators in Coulomb-Gauge Yang-Mills theory employing the path-integral formalism and write down the complete operators for QCD.
We immediately apply them to note that what are usually called the pion square, quartic... charge radii, defined from derivatives of the pion form factor at zero squared momentum transfer, are completely blurred out by relativistic and interaction
corrections, so that it is not clear at all how to interpret these quantities
in terms of the pion charge distribution. The form factor therefore measures matrix elements of powers of the QCD boost and M{\oe}ller operators, weighted by the charge density in the target's rest frame.  
In addition we remark that the decomposition of the $\eta'$ wavefunction
in quarkonium, gluonium, ... components attempted by the KLOE collaboration
combining data from $\phi$ radiative decays, requires corrections due to the
velocity of the final state meson recoiling against a photon. This will be especially
important if such decompositions are to be attempted with data from $J/\psi$ decays.
 % end of PACS codes
} %end of abstract
\authorrunning{G\'omez Rocha, Llanes-Estrada, Sch\"utte and Villalba}
\titlerunning{Coulomb-QCD boosts and applications}
\maketitle

%%%%%%%%%%%%%%%%%%%%%%%%%%%%%%%%%%%%%%%%%%%%%
\section{Introduction}
%%%%%%%%%%%%%%%%%%%%%%%%%%%%%%%%%%%%%%%%%%%%%

%%%%%%%%%%%%%%%%%%%%%%%%%%%%%%%%%%%%%%%%%%%%%
\subsection{The pion form factor}
%%%%%%%%%%%%%%%%%%%%%%%%%%%%%%%%%%%%%%%%%%%%%
Form factors traditionally encode the structure of a composite
target as accessible from elastic reactions. For a scalar target, we
can denote the elastic form factor simply as $\rm  F(Q^2 )$, where
$\rm Q^2=-q^2$, $\rm q=p-p'$. It is conventionally normalized as
$\rm F(0)=\mathcal{Q}$, the particle's charge.

For a non-relativistic particle with unit charge ($\mathcal{Q}=1$, $\rm q^0\simeq 0$),
an expansion around zero momentum transfer allows for a physical
interpretation of the form factor in terms of its rest frame
charge density $\rho\rm (r)$, given by
\begin{equation}
\label{Fexpansion}\rm F(Q^2)= 1-
\frac{1}{3!} \langle r^2\rangle_\rho Q^2 + \frac{1}{5!} \langle
r^4\rangle_\rho Q^4+ {\mathcal O}(Q^6).
\end{equation}  Here we used the notation of
Ref.~\cite{hofstaedter}.  In hadron physics one often quotes also
the curvature of the form factor~\cite{Gasser:1990bv} via
\begin{equation}
\label{Fexpansion2}
\mathrm{F}(\mathrm{Q}^2)=1-\frac{1}{6}\langle \mathrm{r}^2 \rangle \mathrm{Q}^2 + \mathscr{C}_\mathrm{V}^\pi \rm  Q^4 + {\mathcal O}(Q^6).
\end{equation}
Comparing with Eq. (\ref{Fexpansion}) we see that
$\mathscr{C}_\mathrm{V}^\pi = \frac{1}{5!} \langle \mathrm{r}^4 \rangle .$

Beyond the charge normalization
$\int \rm  d^3r \rho(\rm r)=\langle 1 \rangle_\rho = 1$, the derivative at the origin
 provides, in non-relativistic quantum mechanics, the target charge radius
\begin{equation}
\langle \rm r^2 \rangle =-6 \left.\frac{\rm dF}{\rm dQ^2}\right\vert_{\rm Q=0}.
\end{equation} It is known from nuclear physics, but sometimes ignored in
the hadron physics community, that the non-relativistic
interpretation should be modified as effects of boosting the pion
wave function begin to appear. Since the pion is such a light
target, $1/\rm m_\pi^2$ corrections are even larger than the actual
$\langle \rm  r^2 \rangle$ measured experimentally or computed on a
lattice.

The situation is even worse for the curvature, that has been the
object of our recent  focuse.  Using both chiral perturbation theory
and dispersion relations, we have found  a reliable value for the
curvature of $4.0 \pm 0.5\rm \ \ GeV^{-4}$. Many other authors
quoted in \cite{Guo:2008nc} have also dedicated time to extracting
this curvature. It behooves us to examine whether this number --and
the more common $\langle \rm  r^2 \rangle $, have any interpretation
in terms of the target wavefunctions. We show, indeed, that this is
far from trivial.
\begin{figure}
\includegraphics[width=0.45\textwidth,angle=90]{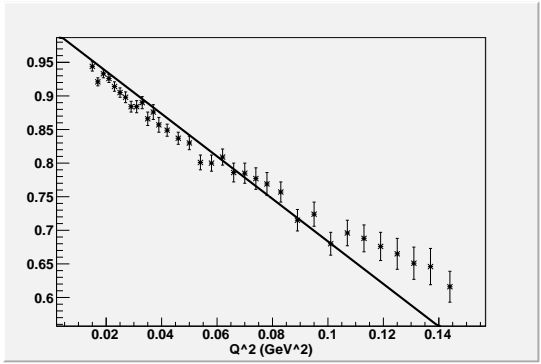}
\parbox[b]{0.52\textwidth}{
\bigskip
\medskip
\includegraphics[width=0.45\textwidth,angle=90]{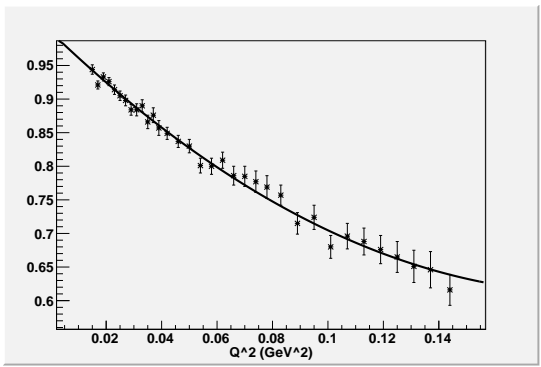}}
\caption{Linear and parabolic fit to the elastic pion form factor at small momentum transfer. The linear fit has $\chi^2/{\rm dof} \simeq 77/32$. The parabolic fit has $\chi^2/{\rm dof} \simeq 22/32$.
\label{fig:pionff}}
\end{figure}
In Fig. \ref{fig:pionff} we replot the form factor data from the NA7
experiment at CERN~\cite{Amendolia:1986wj}, omitting the few points
at higher momentum transfer (as they have large error bars that
might still be underestimated). One can then obtain a direct fit
that is sensitive to the ``square radius'' and curvature. From the
parabolic fit, one can derive $\rm F'(0)=-2.017\pm0.047\ \
GeV^{-2}$, or in units of square radius, $\langle\rm  r^2 \rangle =
0.48 \pm 0.01\ \ fm^2$. The curvature is less well determined,
$\mathscr{C}_\mathrm{v}^\pi\simeq 8 \pm 1\rm \ \ GeV^{-4}$. Adding a
small systematic error to the data the uncertainty in the curvature
could be as high as $2\rm \ \ GeV^{-4}$, still somewhat higher than the
result obtained from the Omn\`es representation quoted above.

Several works in atomic and nuclear physics have pointed out the
necessity of $\rm 1/m$ corrections when talking about the charge
radius of a composite target as extracted from electron scattering
or from the isotope shift~\cite{Huber:1998zz}, for example the so
called Darwin-Foldy terms \cite{Friar:1997js}. In general much of
the literature features idle discussion about where the various
effects are to be adscribed, to the radius, or to its extraction
from electron scattering, or to final state rescattering, etc. In
the end, a lot of the discussion is linguistic. In nuclear physics
one is not quite yet in possession of a full Hamiltonian for the
few-nucleon problem, so the corrections are worked out on a
semiempirical basis \cite{Friar:1979ak} and in atomic physics the
corrections due to both relativity and interactions are small, and
the first order ones can be accessed by means of simple
quantum--mechanical reasoning, although some works exist providing
first--principles computations \cite{Brodsky:1968ea}.

In hadron physics however, the situation is radically different.
Corrections are large, since energies involved match particle masses
(the most salient example being the pion) and interactions strong.
In addition one is in possession of the Hamiltonian of the
fundamental theory,  QCD, and in this article we seek the first
brick in the direction of gaining understanding of form factors in
terms of the canonical quantization of the theory in Coulomb gauge.

Much is known already in light--front \cite{Choi:2008yj} and
point--form quantization \cite{Biernat:2009my,Plessas:2002vq}, with
recent insight from AdS-QCD \cite{Brodsky:2007hb}. However a good
discussion of boost operators of QCD in a phenomenological context
is nowhere to be found in the literature for canonical, Coulomb
gauge quantization. In related quark models there has been indeed
interesting work within the Bakamjian-Thomas construction
\cite{Szczepaniak:1995mi}. Here we attempt to continue stimulating
the discussion and bring it a step closer to QCD.

%%%%%%%%%%%%%%%%%%%%%%%%%%%%%%%%%%%%%%%%%%%%
\subsection{$\phi$-radiative decays and Fock space expansion}
%%%%%%%%%%%%%%%%%%%%%%%%%%%%%%%%%%%%%%%%%%%%

The KLOE collaboration~\cite{DiMicco:2009zza} has
proposed to employ their excellent data set of radiative $\phi$
decays to pseudoscalar mesons, $\phi\to \gamma \eta$, $\phi \to
\gamma \eta'$, to measure the gluonium content of the $\eta'$ as
well as the pseudoscalar mixing angle. The theoretical assumptions
behind the analysis are simple considerations about the flavor
structure of the couplings of gluonium, quarkonium, etc.

While there is no objection to the model analysis of the
pseudoscalar  mixing angle~\cite{Bramon:2000fr}, we believe that a
point has been missed by the community in what regards the gluonium
content of the $\eta'$. This is the fact that the Fock--space
expansion for either of the $\eta$ or $\eta'$
\begin{equation}
\label{Fockexp}
\vert \eta  \rangle = c_1 \vert \mathfrak{q}\bar{\mathfrak{q}}\rangle + c_2 \vert \mathfrak{g}\mathfrak{g} \rangle + c_3 \vert \mathfrak{q}\bar{\mathfrak{q}}\mathfrak{g}\rangle + \dots
\end{equation}
is dependent on the reference frame. When one makes an  assumption
about the gluonium ($\mathfrak{g}\mathfrak{g}$ and larger number of
gluons) content of a meson, one is presumably referring to the
rest-frame of the meson, since the concept of quantum mechanical
state requires a quantization surface that is presumably taken to be
$\rm t=0$ in the rest frame. But, although the $\eta'$ meson is produced
non--relativistically in $\phi$ radiative decays, the analysis
hinges also on decays to the $\eta$ meson, and this is produced with
a velocity $v\simeq 0.55$.

As we will show, even if there was no gluonium at all in either of
the two mesons, one should expect a gluonium content of the $\eta$
meson boosted to such reference frame  of order $\sin \left(\phi_{\rm P}\right)
v^2/2\simeq 0.1$,  given that the pseudoscalar mixing angle seems to
be close to 38 degrees. This is the well-known effect that the boost
operators involve interactions and change the particle content.

This is not necessarily a fatal flaw of the analysis, but a call of
attention that, while effective hadron Lagrangians are explicitly
Lorentz--invariant, such Fock--space decompositions are tied to a
reference frame where canonical QCD quantization is carried out, and
this needs to be specified and consistently handled.

The necessity of specifying a reference frame will become more
accute as the natural next step in the
analysis~\cite{Escribano:2007cd,Escribano:2008rq} will be to employ
the data basis for radiative $J/\psi$ decays accumulated by BES and
others, and given the mass of $3097\ \ \rm MeV$ of the charmonium ground
state, the velocity of the produced mesons will now be decidedly
relativistic. For $J/\psi \to \gamma \eta$, $v_\eta=0.94$, and for
$J/\psi \to \gamma \eta'$, $v_{\eta'}=0.83$, that are indeed
significant.

The rest of this paper is organized as follows. In section~\ref{sec:FF}
we motivate the necessity of constructing the Boost operators with a careful setup of the form factor for a scalar target. Then in subsection~\ref{subsec:pion} we particularize to the case of the pion, the lightest hadron where relativistic corrections are most significant. Subsection~\ref{subsec:Breit} is then dedicated to dispelling some misconception about the charge radius interpreted in the Breit frame.

Although pieces thereof can be found in the literature, it is convenient to write down the boost operators of QCD in Coulomb gauge for future reference, which we do in section~\ref{sec:Poincare}. There we also present a relatively easy derivation based on modern functional methods in subsections~\ref{subsec:path},\ref{subsec:poincare} and \ref{subsec:quark}.

Finally, section~\ref{sec:phidec} presents the application of these boost operators to $\phi\to \eta \gamma$ (or similar decays) where it has been until now ignored, and section~\ref{sec:conclusions} presents our final remarks.

%%%%%%%%%%%%%%%%%%%%%%%%%%%%%%%%%%%%%%%%%%%%%
\section{Rigorous interpretation of the form factor in canonically quantized field theory} \label{sec:FF}
%%%%%%%%%%%%%%%%%%%%%%%%%%%%%%%%%%%%%%%%%%%%%
The form factor can be expressed, in non--relativistic normalization for the charge density, as
\begin{equation}
_{\rm out}\langle \textrm{p}'\vert  \rm j^\mu \vert \textrm{p}
\rangle_{\rm in} =
\frac{(\textrm{p}+\textrm{p}')^\mu}{2\sqrt{\textrm{EE}'}} \rm
F(q^2).
\end{equation}
The energy-momentum $\rm p^\mu=(m_\pi, {\bf 0})$ corresponds to a pion
at rest and time $\rm t=-\infty$ before the momentum transfer by the
virtual photon, $\rm q^2$, boosts it to a frame with $\rm
p^{\prime}$.

The charge density is taken at $\rm t=0$, that we employ as
quantization surface to define the equal-time commutation and
anticommutation relations for field operators. To undertake any
computation one needs to propagate the initial state $\vert \rm p
\rangle_{\rm in}$ to $\rm t=0$. The propagators that accomplish this
operation are usually called Moeller operators in the interaction
picture, $\Omega_-=\rm U(t=-\infty,0)$. Likewise the propagation of
the final state to the quantization surface is carried out by
$\Omega_+^\dagger=\rm U^\dagger (t=0,\infty)$. The product of both
operators would reconstruct the $\textsc{S}$ matrix,
$\Omega_+^\dagger\Omega_-=\textsc{S}$, but here an insertion of the
current operator occurs at time $\rm t=0$, see
Eq.~(\ref{eq:formfactor}) below.

The final--state pion is boosted, and to be able to use its
wavefunction  in the rest frame (where the charge density is
defined) one needs to employ the boost $\hat{\mathbf{K}}$ operator
from QCD
\begin{equation} \label{defboost}
_{\mathrm{in}}\langle\mathrm{p}^{\prime} \vert= _{\mathrm{in}}\langle\mathrm{p} \vert
\mathrm{e}^{-i\hat{\mathbf{K}}\cdot\boldsymbol{\zeta}}
\end{equation}
that we construct in Sec. \ref{sec:Poincare} below.
In terms of these operators the form factor can be expressed via
\begin{equation} \label{eq:formfactor}
_{\mathrm{in}}\langle \mathrm{p} \vert \mathrm{e}^{-i\hat{\mathbf{K}}\cdot\boldsymbol{\zeta}}
\Omega_+^\dagger j^\mu(0) \Omega_- \vert \rm p\rangle_{\rm in} =
\frac{\rm (p+p')^\mu}{2\sqrt{\rm EE'}}\rm  F(q^2)
\end{equation}
(the M{\oe}ller operators introduce interactions  both sides of the
current insertion). An alternative form of $\Omega_+^\dagger \rm
j^\mu(0) \Omega_-$ is for example $\textrm{T}\left(\textrm{j}^\mu(0)
\textrm{e}^{i \int \textrm{d}^4
\textrm{x}\mathfrak{L}_{\textrm{I}}}\right).$

The interpretation of the form factor in non-relativistic quantum mechanics
(where it comes about in the Born approximation by expanding the Fourier transform of the potential causing the scattering)
\begin{equation}\label{FFnorel}
\rm F_{\mathrm{NonRel}}(\vert\mathbf{q}\vert^2) = \mathcal{Q} - \frac{1}{3!}
\vert\textbf{q}\vert^2 \langle r^2 \rangle +
\frac{1}{5!}\vert\textbf{q}\vert^4\langle r^4 \rangle
\end{equation}
in terms of charge radii, is only a limiting case that can be recovered from Eq. (\ref{eq:formfactor}) and we will do it shortly.

Let us now obtain an equivalent $\rm q^2/m_\pi^2$ expansion of
Eq.~(\ref{eq:formfactor}) in powers of the momentum transfer to
match Eq.~(\ref{FFnorel}).  The
rapidity parameter in Eq.~(\ref{defboost}) depends on the
transferred momentum through
\begin{equation}
v= \tanh{\zeta}\ \ \Rightarrow \ \    \zeta = \frac{1}{2}\log \left( \frac{1+v}{1-v}\right) \label{rapidity}
\end{equation} From the above equation we can approximate
\begin{equation}
\zeta \simeq v + \frac{v^3}{3}+\dots O (v^5) \ \ \mathrm{with}\ \
v^2 \simeq -\frac{\rm q^2}{\rm m_\pi^2} - \frac{3}{4}\frac{\rm q^4}{\rm
m_\pi^4}.
\end{equation}
Expanding also $\rm E'$ in terms of $\rm q$, and making ${\bf
q}\propto\bf n_z$  ($\bf n_z=k_z/\vert k_z\vert$) infinitesimal so
that one can truncate the Taylor expansion of the boost
\begin{equation}
\exp\left(i \hat{\rm K}_z\zeta_z\right)=  1 +i\hat{\rm K}_z\zeta_z
-\frac{1}{2}\hat{\rm K}^2_z \zeta_z^2 - \frac{i}{3!}\hat{\rm K}^3_z
\zeta_z^3+\frac{1}{4!}\hat{\rm K}^4_z \zeta_z^4+\ldots
\end{equation}
we obtain
\begin{eqnarray}
\rm F(q^2) &=& \frac{2\sqrt{\rm EE'}}{(\rm p+p')^0}
\langle \rm p \vert e^{-i\hat{\rm K}_z\zeta_z} \Omega_+^\dagger \rm
j^0(0) \Omega_- \vert\rm  p\rangle\\ \nonumber &\simeq& \left( 1+
\frac{\rm q^2}{2\rm m_\pi^2}+ \frac{\rm q^4}{32\rm m_\pi^4}\right)\times \nonumber\\
&\times&\langle \rm p \vert\left( 1+
\frac{\hat{K}_z^2}{2m_\pi^2}q^2+\frac{\hat{K}_z^2}{24m_\pi^4}q^4+\frac{\hat{K}_z^4}{4!m_\pi^4}q^4\right)
\;\Omega_+^\dagger\; j^0 \Omega_- \vert p\rangle \nonumber
\end{eqnarray}
(terms odd in $\hat{\rm K}_z$ are absent for a scalar target because
of $z\to -z$ reflection symmetry). We then have
\begin{equation}
\rm F(q^2)  \simeq \rm  F(0)+\rm q^2 F^\prime (0) + \frac{1}{2!}q^4
F^{\prime\prime}(0)+ \dots
\end{equation}
where the derivatives are now
\begin{eqnarray} \label{relcorr1}
\begin{array}{c}
\rm F(0)=\langle p | \Omega_+^\dagger j^0\Omega_- \vert p\rangle, \\  \\  \rm F'(0)=\frac{1}{2m_\pi^2}F(0) + \frac{1}{2m_\pi^2} \langle p | \hat{K}_z^2
\Omega_+^\dagger j^0\Omega_- \vert p\rangle \\ \\ 
\rm F''(0) = \frac{1}{12m_\pi^4}\langle p | \hat{K}_z^4 \Omega_+^\dagger
j^0\Omega_- \vert p\rangle - \frac{25}{48
m_\pi^4}F(0)+\frac{7}{6m_\pi^2}F'(0).
\end{array}
\end{eqnarray}

%%%%%%%%%%%%%%%%%%%%%%%%%%%%%%%%%%%%%%%%%%%%%%%%%%%%%%%%%%%%%%%
\subsection{No square radius interpretation for the pion} \label{subsec:pion}
%%%%%%%%%%%%%%%%%%%%%%%%%%%%%%%%%%%%%%%%%%%%%%%%%%%%%%%%%%%%%%%%

To reproduce Eq.~(\ref{FFnorel}), one first invokes the  impulse
approximation, by which the quarks from the $\rm j^\mu$ current
density directly enter the state's wavefunction, in effect
neglecting the M{\oe}ller operators (rescattering). This is known from
nuclear physics to be a poor approximation, and many works in hadron
physics have also lifted it~\cite{Maris:2000sk} since it fails, for
example, to reproduce simple instances of Vector Meson Dominance.

In a second step, one takes the non-relativistic limit by sending
the mass denominators $\rm m_\pi\to\infty$, and substitutes the Lorentz
boost by its Galilean equivalent. Since a Galilean change of
reference frame  is generated by
\begin{equation}
\mathrm{U}[\mathbf{v}]=\exp\left(-i \rm
m_\pi\hat{\mathbf{r}}\cdot\mathbf{v}\right),
\end{equation} the relevant boost
operator in quantum mechanics becomes $ \hat{\mathbf{K}} \rightarrow
\hat{\mathbf{K}}_{\mathrm{Galilean}}=\rm -m_\pi\hat{\mathbf{r}}$.
Upgrading it to non-relativistic field theory for quarks, and
summing over spin, color and flavor, the boost operator is then
\begin{equation}
\hat{\mathbf{K}}_{\mathrm{Galilean}}= \sum_\ell \int \rm d^3 x \;
\hat{\mathfrak{q}}^{\ell\dagger}(x)
(-m_{\mathfrak{q}_\ell}\mathbf{x}) \hat{\mathfrak{q}}^\ell (x).
\end{equation}
(In terms of quadrispinors including antiparticles, there would be
an additional Dirac beta matrix $(-i\rm m_{\mathfrak{q}_j}) \beta
 \partial/\partial \rm k^i$ in the operator).

In this case, one obtains for the form factor a non-relativistic
expression in terms of the wavefunction of the target. If the target
$\vert\rm  H \rangle$ is taken as a two-body scalar with
non-relativistic wavefunction $\rm \mathscr{F}(r)$, and two particles of mass
$\rm m_1$, $\rm m_2$ in the center of mass frame, with relative
position $\textbf{r}$, and opposite charge (as corresponds for
example to the non-relativistic Hydrogen atom or the neutral $\pi_0$
in the quark model)
\begin{equation}
\langle \rm r^2 \rangle = \rm e\int d^3 x r^2\vert \mathscr{F}(r)\vert^2
\frac{m_1-m_2}{m_1+m_2}.
\end{equation}

However, returning to Eq.~(\ref{relcorr1}) for the case of the pion,
$1/2\rm m_\pi^2\simeq 26\ \ GeV^{-2}$ is not negligible in any sense
against $-\rm F'(0) \simeq 2\ \ GeV^{-2}$ taken from the
experimental data quoted above. There need to be large cancellations
among the two terms in the  second of Eq.~(\ref{relcorr1}). Its
second non-constant term, that in the non-relativistic limit would
correspond to the square radius, is therefore of order $20\ \ \rm
GeV^{-2}$. One needs to conclude that, in the case of the pion,
there is no connection whatsoever between the derivative of the
spacelike form factor and the square radius of the rest frame charge
distribution.

\begin{figure}
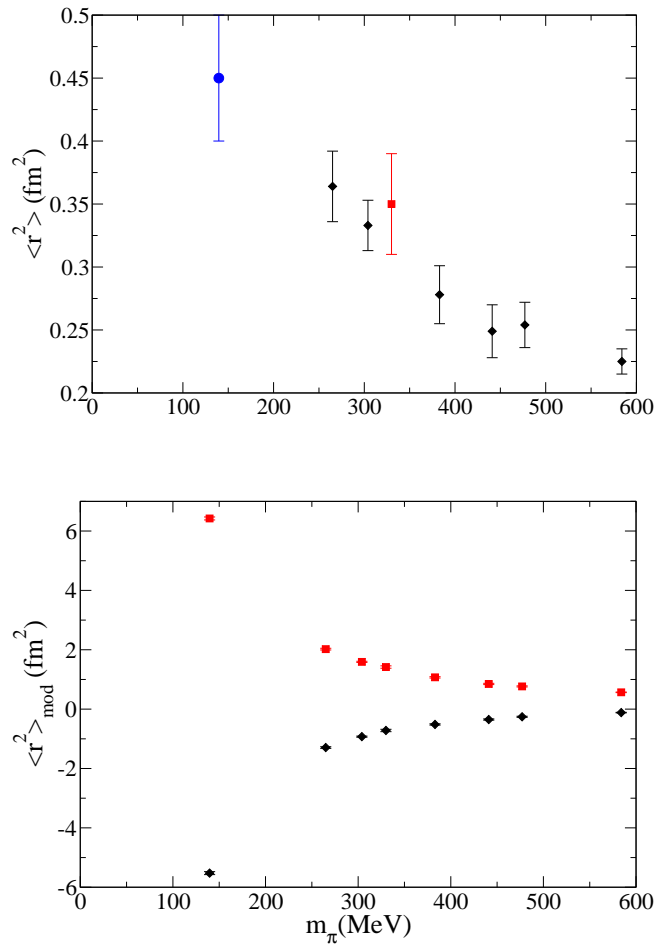

\includegraphics[width=0.47\textwidth]{latticersq.eps}
\parbox[b]{0.52\textwidth}{
\bigskip
\medskip
\includegraphics[width=0.47\textwidth]{latticemod2.eps}}
\caption{Top panel: accepted~\cite{Amsler:2008zzb} value of $-6\rm
F'(0)$, the would-be pion square charge radius,  and lattice
calculations \cite{latticepapers} as a function of the pion mass.
Bottom panel: the two terms of Eq.~(\ref{relcorr1}), that are way
larger, and a cancelation must occur to give the physical form
factor derivative. It is not clear that one can extract the square
charge radius, except for large quark masses where one could think
of the lattice data in non--relativistic terms.
\label{fig:latticedata}}
\end{figure}

%%%%%%%%%%%%%%%%%%%%%%%%%%%%%%%%%%%%%%%%%%%%%%%%%%%%%%%%%%%%%
\subsection{Breit frame}\label{subsec:Breit}
%%%%%%%%%%%%%%%%%%%%%%%%%%%%%%%%%%%%%%%%%%%%%%%%%%%%%%%%%%%%%
In the study of the proton form factor, the Breit frame became very
popular \cite{Sachs:1962zzc,Kelly:2002if}. One reason is that, for a
spin $1/2$ target, a separation of the charge and magnetization
densities is possible by employing the Sachs form factors. The point
that concerns us here, the pion being a scalar target, is that the
$1\rm /m_\pi$ terms in Eq.~(\ref{relcorr1}) for $\rm F'(0)$ are
absent.

The Breit (or brick--wall) frame is defined as that in which the
target bounces off the virtual photon with opposite incoming and
outgoing momentum. As a consequence, $\rm E_{\rm Breit}=E'_{\rm
Breit}$ and the factor  $\rm (p+p')^0/2\left(\rm EE'\right)^{1/2} $ in
Eq.~(\ref{eq:formfactor}) and following becomes unity.

Therefore in this frame,
\begin{equation}
\rm F(q^2) = \left\langle + \left.\frac{p}{2} \right\vert
\Omega^\dagger_+ j^0(0) \Omega_- \left\vert - \frac{p}{2}
\right.\right\rangle
\end{equation}
and therefore
\begin{eqnarray} \label{relcorr2}
\begin{array}{c}
\rm F(0)=\left\langle - \frac{p}{2} \right\vert \Omega_+^\dagger
j^0\Omega_- \left\vert - \frac{p}{2}\right\rangle, \\ \\
\rm F'(0)=\frac{1}{2m_\pi^2} \left\langle  - \frac{p}{2} \right\vert
\hat{K}_z^2 \Omega_+^\dagger j^0\Omega_- \left\vert -
\frac{p}{2}\right\rangle,\\ \\
\rm F''(0) = \frac{1}{12\rm m_\pi^4}\left\langle  - \frac{\rm p}{2}
\right\vert \hat{\rm K}_z^4 \Omega_+^\dagger\rm  j^0\Omega_-
\left\vert - \frac{\rm p}{2}\right\rangle +\frac{1}{6\rm m_\pi^2}\rm
F'(0).
\end{array}
\end{eqnarray} 

In principle, the large term in $\rm F'(0)$ proportional to $\rm
F(0)$ in Eq.~(\ref{relcorr1}) is absent, although the interpretation
of the remaining in terms of a square charge radius is still not
possible due to the interactions (M{\oe}ller operators) and the
necessary change of momentum (and hence frame) of the incoming and
outgoing target. Moreover, the curvature in $\rm F''(0)$ is still
affected by $1/\rm m_\pi^2$ corrections. Therefore there is no clear
advantage of interpretation of the lattice or experimental data in
employing the Breit frame, at least in the case of the pion.

%%%%%%%%%%%%%%%%%%%%%%%%%%%%%%%%%%%%%%%%%%%%%
\section{Poincar\'e generators of QCD \label{sec:Poincare}}
%%%%%%%%%%%%%%%%%%%%%%%%%%%%%%%%%%%%%%%%%%%%%

Having motivated their phenomenological importance, in this section
we derive the generators of the Poincar\'e transformations in
Coulomb gauge QCD. The method we follow is a modern path-integral
approach in the first--order formalism  (with the path integration
extending over phase space $(\rm q,p)$ \cite{VillalbaChavez:2008dv},
not only the field coordinates $\rm q$), based on the work of
Zwanziger \cite{Zwanziger:1998ez} and Watson and Reinhardt \cite{Watson:2006yq}. First we obtain the
result for Yang-Mills theory in the functional approach. Then we
derive the result for the boost generators appropriate for canonical
quantization, in agreement with~\cite{Besting:1989nq}. Finally we
add the quarks to complete the Boost operators in canonically
quantized QCD.

%%%%%%%%%%%%%%%%%%%%%%%%%%%%%%%%%%%%%%%%%%%%%%%%%%%%%%%%%%%%%%%%%%%%%%%%%%%%%%%%%%%
\subsection{The Hamiltonian and Momentum Generators of a pure Yang-Mills theory}\label{subsec:path}
%%%%%%%%%%%%%%%%%%%%%%%%%%%%%%%%%%%%%%%%%%%%%%%%%%%%%%%%%%%%%%%%%%%%%%%%%%%%%%%%%%%

Yang-Mills theory is based on the renormalizable Lagrangian density
of a spin-1, color--octet field $\rm A_\mu^a(\rm x)$, and written
down in a gauge invariant manner in terms of the Maxwell tensor
$\mathfrak{F}_{\mu\nu}^a=\partial_{\mu}\rm
A_{\nu}^a-\partial_{\nu}A_{\mu}^a +gf^{abc}A_{\mu}^bA_{\nu}^c$ as
\begin{equation}
\mathcal{S}_{\mathrm{YM}}=\int \rm d^4x\mathscr{L}_{\mathrm{YM}}\ \
\mathrm{with} \ \
\mathscr{L}_{\mathrm{YM}}=-\frac{1}{4}\mathfrak{F}_{\mu\nu}^a\mathfrak{F}^{a\mu\nu}\
. \label{LagrangianYM}
\end{equation}
The canonical stress-energy tensor conserved by translational invariance of
$\mathscr{L}_{\mathrm{YM}}$ reads
\begin{eqnarray}
\mathscr{T}^{\mu\nu}_\mathrm{YM}&=&\frac{\partial\mathscr{L}_{\mathrm{YM}}}{\partial(\partial_\mu\mathrm{A}_\lambda^{a})}\partial^\nu\mathrm{A}_{\lambda}^{a}-g^{\mu\nu}\mathscr{L}_{\mathrm{YM}}\\\nonumber\\&=&-\mathfrak{F}^{\mu a}_{\ \ \lambda}\partial^{\nu}\mathrm{A}^{\lambda a}+\frac{1}{4}g^{\mu\nu}\mathfrak{F}^{\sigma a}_{\lambda}\mathfrak{F}^{\sigma a}_{\lambda}\ , \label{gaugesdlsdiif}
\end{eqnarray}
$g^{\mu\nu}$being  the Minkowski metric with signature $+---$.
Associated to space-time translational invariance are the conserved
Hamiltonian and Momentum
\begin{equation}
\textrm{H}_{\mathrm{YM}}=\int \rm d^3x\mathscr{T}_{\mathrm{YM}}^{00} \ \
\mathrm{and}\ \ \boldsymbol{\mathcal{P}}_{\mathrm{YM}}=\int \rm d^3x\mathscr{T}_{\mathrm{YM}}^{0i}.\label{gjdoa}
\end{equation}

In the path-integral framework the Green's functions are generated from the functional
\begin{equation}
\mathcal{Z}=\int\mathcal{D}\Phi\exp\left\{
i\mathcal{S}_\mathrm{YM}\right\}
\end{equation} 
($\Phi$ representing
generically an integral over all fields). Since we are interested in
exposing the explicit transverse gluons, the degrees of freedom of
the intuitive Coulomb gauge, we divide the Faraday-Maxwell tensor in
terms of chromoelectric and chromomagnetic fields
\begin{eqnarray}
\mathbf{E}^a=-\partial^0\mathbf{A}^a-\boldsymbol{\nabla}\mathrm{A}^{0a}+\mathrm{gf}^{abc}\mathbf{A}^b\mathrm{A}^{0c}\\
\nonumber
\mathrm{B}_i^a=\epsilon_{ijk}\left[\nabla_j\mathrm{A}_k^a-\frac{1}{2}
\mathrm{gf}^{abc}\mathrm{A}_j^b\mathrm{A}_k^c\right].
\end{eqnarray} Coulomb gauge is fixed with the
help of the Faddeev-Popov mechanism by first introducing a
gauge--fixing term in the action
\begin{eqnarray}
\label{Zfunctional}
\begin{array}{c}
\mathcal{Z}=\int\mathcal{D}\Phi\exp\left\{i\mathcal{S}_{\mathrm{YM}}+i\mathcal{S}_{\mathrm{FP}}\right\},\\
\\ \mathcal{S}_{\mathrm{FP}}=\int \mathrm{d}^3\mathrm{x}\left[-\xi^a\boldsymbol{\nabla}\cdot\mathbf{A}^a-\bar{c}^a\boldsymbol{\nabla}\cdot \mathbf{D}^{ab}c^b\right].
\end{array}
\end{eqnarray} Here $c^a$ are Grassmann (ghost) fields that keep track of the
Faddeev-Popov determinant for now, involving the
covariant-derivative operator  $\mathbf{D}^{ab}=
\delta^{ab}\boldsymbol{\nabla} + \mathrm{g f}^{abc}\mathbf{A}^c$ and
$\xi^a$ is a Lagrange multiplier to assist with the Coulomb gauge
constraint.

As is known, Coulomb gauge fixing is not unique in non-Abelian gauge
theories. Gribov \cite{gribov} was the first to point out that there
remain  physically equivalent gauge  configurations that are related
by finite (as opposed to infinitesimal) gauge transformations. A
further restriction must be imposed to the configuration space of
gauge fields to a region  where
\begin{equation} \label{Coulombrest}
\Omega\equiv\left\{\mathbf{A}:\ \ \boldsymbol{\nabla}\cdot\textbf{A}=0\ \ \vert\ \ -\boldsymbol{\nabla}\cdot\textbf{D}\geq0\right\}.
\end{equation}This simple domain is still not totally free of Gribov copies. Instead one should consider  the fundamental modular region $\Lambda$ \cite{zwanziger2}. It turns out though that functional integrals are dominated by configurations on the common boundary of $\Omega$ and  $\Lambda$  \cite{Watson:2006yq} so that, in practice, it is enough to consider the domain defined by Eq. (\ref{Coulombrest}). 

The canonical momentum associated to the transverse $\mathbf{A}^a$
fields is introduced as an auxiliary field thanks to the following
identity
\begin{eqnarray}
\exp{\left\{i\int \rm
d^4x\frac{1}{2}\mathbf{E}^a\cdot\mathbf{E}^a\right\}} =\nonumber\\
\int\mathcal{D}\boldsymbol{\pi}\exp{\left\{i\int \mathrm{d}^4\mathrm{x}\left[-\frac{1}{2}
\boldsymbol{\pi}^a\cdot \boldsymbol{\pi}^a-\boldsymbol{\pi}^a\cdot
\mathbf{E}^a\right]\right\}}.
\end{eqnarray} Two more auxiliary variables $\Omega$, $\tau$ will allow
dividing the chromoelectric part of the action into the dynamical
transverse  potential of Coulomb gauge and a constrained
longitudinal part. One needs the further identity
\begin{eqnarray}
\mathrm{const}&=&\int\mathcal{D}\Omega\delta\left(\boldsymbol{\nabla}\cdot
\boldsymbol{\pi} +\nabla^2\Omega\right)\nonumber \\ \nonumber
&=&\int\mathcal{D}\left\{\Omega,\tau\right\}\exp{\left\{-i\int \mathrm{d}^4\mathrm{x}\tau^a\left(\boldsymbol{\nabla}\cdot\boldsymbol{\pi}^a+\nabla^2\Omega^a\right)\right\}}.
\end{eqnarray} and a change of variables $\boldsymbol{\pi}\rightarrow
\boldsymbol{\pi}-\boldsymbol{\nabla}\Omega $ to arrive to the equivalent form of
the Yang-Mills action
\begin{eqnarray}
\mathcal{S}_{\mathrm{YM}}=\int \mathrm{d}^4\mathrm{x}
\left[-\frac{1}{2}\textbf{B}^a\cdot\textbf{B}^a-\tau^a\boldsymbol{\nabla}\cdot\boldsymbol{\pi}^a
\right.\nonumber\\-\left.\frac{1}{2}(\boldsymbol{\pi}^a-\boldsymbol{\nabla}\Omega^a)\cdot(\boldsymbol{\pi}^a-\boldsymbol{\nabla}\Omega^a)
\right.\nonumber  \\ \left.
+(\boldsymbol{\pi}^a-\boldsymbol{\nabla}\Omega^a)\cdot\left(\partial^0\mathbf{A}^a+\mathbf{D}^{ab}\mathrm{A}^{0b}
\right)\right].
\end{eqnarray}

We now perform the integral over the ghost fields in Eq.
(\ref{Zfunctional}) to recover the Jacobian of the change of
variables. Likewise we integrate  over the Lagrange multipliers
$\xi^a$  and $\tau^a$ to recover the delta-functions associated to
the constraints, that enforce transversality of the physical fields
(the Coulomb gauge condition $\boldsymbol{\nabla}\cdot \mathbf{A}=
\boldsymbol{\nabla}\cdot \boldsymbol{\pi}=0$). Thereafter one can set to zero any
terms involving these divergences in the action.

Taking these details into account $\mathcal{Z}$ can be expressed as
\begin{eqnarray}
\label{Zintermediate}
\mathcal{Z}=\int\mathcal{D}\Phi\mathrm{Det}\left[-\boldsymbol{\nabla}\cdot\mathbf{D}\delta^4(\rm
x-y)\right]
\delta\left(\boldsymbol{\nabla}\cdot\mathbf{A}\right)\\\times\delta\left(\boldsymbol{\nabla}\cdot\boldsymbol{\pi}\right)
\exp{\left\{i\mathcal{S}\right\}}\nonumber
\end{eqnarray}with
\begin{eqnarray}
\mathcal{S}&=&\int \mathrm{d}^4\mathrm{x}\left[-\frac{1}{2}\textbf{B}^a\cdot\textbf{B}^a
-\frac{1}{2}\boldsymbol{\pi}^{a}\cdot\boldsymbol{\pi}^a+\frac{1}{2}\Omega^{a}\nabla^2\Omega^a\right.\nonumber\\
&+&\left.\boldsymbol{\pi}^a\cdot\partial^0\mathbf{A}^a+\mathrm{A}^{0a}\left(\boldsymbol{\nabla}\cdot\mathbf{D}^{ab}\Omega^b
+\mathrm{g}\varrho_\mathfrak{g}^a\right)\right]
\end{eqnarray}and
$\varrho_\mathfrak{g}^a=\mathrm{f}^{ade}\mathbf{A}^d\cdot\boldsymbol{\pi}^e$
the color--charge density carried by the gluons.

The functional integral over $\rm A^0$ can also be performed to
yield the constraint equivalent to Poisson's equation in Quantum
Electrodynamics in the form of a new delta--function on the path
integral, $\delta\left(-
\boldsymbol{\nabla}\cdot\mathbf{D}^{ab}\Omega^{b}-\mathrm
{g}\varrho_\mathfrak{g}^a\right)$. The Coulomb instantaneous potential
of QED, $1/\vert \textbf{x}-\textbf{y}\vert$, inverse of the
Laplacian, is generalized in QCD to the inverse of the Faddeev-Popov
operator
\begin{equation}
\left[-\boldsymbol{\nabla}\cdot\mathbf{D}^{ab}\right] \mathrm{M}^{bc}=\delta^{ac},
\end{equation}
and the constraint can be formally solved for $\Omega$, now taking
the place of the scalar potential $\rm A^0$,
\begin{equation}
\Omega^a=
\mathrm{gM}^{ab}\varrho_\mathfrak{g}^b\label{omesoliuti}
\end{equation}
The associated delta--function factorizes in a useful way~\cite{Reinhardt:2008pr}
\begin{eqnarray}
 \label{FPappears}
\delta\left(-\boldsymbol{\nabla}\cdot\mathbf{D}^{ab}\Omega^b-\mathrm{g}\varrho_\mathfrak{g}^a\right)\nonumber\\
= \mathrm{Det}\left[-\boldsymbol{\nabla}\cdot\mathbf{D}\delta^4(\rm
x-y)\right]^{-1}
\delta\left(\Omega^a-\mathrm{g}\mathrm{M}^{ab}\varrho_\mathfrak{g}^b\right)
\end{eqnarray} to cancel the determinant in Eq. (\ref{Zintermediate}) yielding
\begin{equation}
\mathcal{Z}=\int\mathcal{D}\mathbf{A}\mathcal{D}\boldsymbol{\pi}\delta\left(\boldsymbol{\nabla}\cdot\mathbf{A}\right)
\delta\left(\boldsymbol{\nabla}\cdot\boldsymbol{\pi}\right)\exp{\left\{i\mathcal{S}_0\right\}}\label{finalpathintergralrepre}
\end{equation} where
\begin{eqnarray}
\mathcal{S}_0&=&\int 
\mathrm{d}^4\mathrm{x}\left[-\frac{1}{2}\textbf{B}^a\cdot\textbf{B}^a -\frac{1}{2}
\boldsymbol{\pi}^{a}\cdot\boldsymbol{\pi}^a\right.\\ &-&\left.\frac{1}{2}
\mathrm{g}^2\varrho_\mathfrak{g}^{b}\mathrm{M}^{ba}(-\nabla^2)\mathrm{M}^{ac}\varrho_\mathfrak{g}^c
+\boldsymbol{\pi}^a\cdot\dot{\mathbf{A}}^a\right].
\end{eqnarray}

Clearly,  the term  $ \boldsymbol{\pi}^a\cdot \dot{\mathbf{A}}^a $ is
equivalent to $\rm p\dot{q}$ in classical mechanics, and the
remaining part can be identified  as the classical  Hamiltonian of
pure Yang-Mills theory
\begin{eqnarray}
\begin{array}{c} \displaystyle
\rm H_{\mathrm{YM}}= \int\mathscr{H}_{\mathrm{YM}}(\textbf{x},\rm
t)\rm d^3x  \\ \\  \displaystyle
\mathscr{H}_{\mathrm{YM}}=\frac{1}{2}\boldsymbol{\pi}^{a}(\textbf{x},\mathrm
{t})\cdot\boldsymbol{\pi}^a(\textbf{x},\mathrm
{t})+\frac{1}{2}\textbf{B}^a(\textbf{x},\mathrm
{t})\cdot\textbf{B}^a(\textbf{x},\mathrm{t})\\ \\ +\displaystyle
\frac{1}{2}\mathrm{g}^2\varrho_\mathfrak{g}^{b}\mathrm{M}^{ba}(-\nabla^2)\mathrm{M}^{ac}\varrho_\mathfrak{g}^c. 
\end{array}
\label{YMHamiltonia12n1d}
\end{eqnarray}

We  now take our expressions to the canonical functional
quantization formalism.  At this point $\boldsymbol{\pi}$ ceases to be  an
integration variable and  must be considered as the standard
classical  canonical momentum $\boldsymbol{\pi}^a=\partial
\mathscr{L}_{\mathrm{YM}}/\partial(\dot{\mathbf{A}})^a.$ After canonical quantization, the transverse field $\boldsymbol{\pi}^a$  has to be substituted by the momentum conjugate $\hat{\mathbf{\Pi}}^a$  to the transverse field $\hat{\mathbf{A}}^{a}.$ This is however not trivial, but the problem has already been solved in the past.

Indeed, comparing this Hamiltonian with that of Christ and Lee
\cite{Christ:1980ku}, or in recent papers
\cite{Feuchter:2004mk,Szczepaniak:2001rg,Szczepaniak:2003ve,Campagnari:2009km},
one notices the absence of the Faddeev-Popov determinant  in our
expression. In order to introduce
$\mathcal{J}[\mathbf{A}]=\mathrm{Det}(-\boldsymbol{\nabla}\cdot\hat{\mathbf{D}})$  we first note that  the path integral representation in Eq.
(\ref{finalpathintergralrepre}) involves a Cartesian integration measure over gauge field.  Let us consider, then, the corresponding  quantized version of  Eq.
(\ref{YMHamiltonia12n1d}) 
\begin{eqnarray}\label{shethamivaria}
\hat{\tilde{\mathscr{H}}}_{\mathrm{YM}}&=&\frac{1}{2}\hat{\tilde{\mathbf{\Pi}}}^{a\dagger}(\textbf{x})\cdot\hat{\tilde{\mathbf{\Pi}}}^a(\textbf{x})+\frac{1}{2}\hat{\textbf{B}}^a(\textbf{x})\cdot\hat{\textbf{B}}^a(\textbf{x})\\&+&\frac{1}{2}
\mathrm{g}^2\mathrm{f}^{bde}\hat{\mathbf{A}}^d(\textbf{x})\cdot\hat{\tilde{\mathbf{\Pi}}}^{e\dagger}(\textbf{x})\hat{\mathrm{M}}^{ba}(-\nabla^2)\hat{\mathrm{M}}^{ac}\hat{\tilde{\varrho}}_\mathfrak{g}^c\nonumber
\end{eqnarray}

One is not entitled to use these operators (that we have covered with a tilde)  as the canonical momentum since $ \hat{\tilde{\mathbf{\Pi}}}^a \neq -i\delta/\delta \hat{\mathbf{A}}^a(\mathbf{x})$ (we repair this below in Eq. (\ref{moemntumtransformationseps})). In a Hilbert space  where  the scalar product of the complex-valued
wavefunctionals $\tilde{\Psi}[\mathbf{A}]$ is  taken as ``Cartesian'':
\begin{equation}
\langle\tilde{\Psi}_1\vert\tilde{\Psi}_2\rangle=\int\mathcal{D}\mathbf{A}\tilde{\Psi}_1^*[\mathbf{A}]\tilde{\Psi}_2[\mathbf{A}]\ . \label{pasandoaJvar}
\end{equation} (Note that in Eq. (\ref{shethamivaria}) we have fixed $\rm t=0$).

We now, introduce
\begin{equation}
\label{redefPsi}
\Psi_i[\mathbf{A}]\equiv\mathcal{J}^{-1/2}[\mathbf{A}]\tilde{\Psi}_i[\mathbf{A}],
\end{equation}which is the wavefunctional one would find if,
as Christ and Lee did, one had first quantized in $\rm A^0=0$ gauge
and then  transformed the gauge fields to Coulomb gauge, since in
this variable change the scalar product picks up the determinant Eq.
(\ref{pasandoaJvar})
\begin{equation}
\langle\Psi_1\vert\Psi_2\rangle=\int\mathcal{D}\mathbf{A}\mathcal{J}[\mathbf{A}]\Psi_1^*[\mathbf{A}]\Psi_2[\mathbf{A}] \label{pasandoaJ2}
\end{equation}
that we would have instead absorbed in the wavefunctional.  However, by doing the  change of wave   functional Eq. (\ref{redefPsi}) one should  at the same time preserve the expectation value of observables 
$\langle\tilde{\Psi}_1\vert\hat{\tilde{\mathcal{O}}}\vert\tilde{\Psi}_2\rangle=\langle\Psi_1\vert\hat{\mathcal{O}}\vert\Psi_2\rangle$ with 
\begin{equation}
\hat{\mathcal{O}}=\mathcal{J}^{-1/2}\hat{\tilde{\mathcal{O}}} \mathcal{J}^{1/2}.
\end{equation}This has to be true, in particular for the Hamiltonian density. As a consequence
\begin{eqnarray}
\begin{array}{c}
\hat{\mathscr{H}}=\mathcal{J}^{-1/2}\hat{\tilde{\mathscr{H}}} \mathcal{J}^{1/2}\\ \\
\hat{\mathscr{H}}=\frac{1}{2}\mathcal{J}^{-1/2}\hat{\tilde{\mathbf{\Pi}}}^{a\dagger}(\textbf{x})\hat{\tilde{\mathbf{\Pi}}}^a(\textbf{x})\mathcal{J}^{1/2}+\frac{1}{2}\hat{\textbf{B}}^a(\textbf{x})\cdot\hat{\textbf{B}}^a(\textbf{x})\\ \\ +\frac{1}{2}
\mathrm{g}^2\mathcal{J}^{-1/2}\hat{\tilde{\varrho}}_\mathfrak{g}^{b}\hat{\mathrm{M}}^{ba}(-\nabla^2)\hat{\mathrm{M}}^{ac}\hat{\tilde{\varrho}}_\mathfrak{g}^c\mathcal{J}^{1/2}.
\end{array}
\end{eqnarray}
Replacing the operator $\hat{\tilde{\mathbf{\Pi}}}^a(\textbf{x})$ by the transformed one
\begin{eqnarray}\label{moemntumtransformationseps}
\hat{\tilde{\mathbf{\Pi}}}^a(\textbf{x}) &=&\mathcal{J}^{1/2}\hat{\mathbf{\Pi}}^a(\textbf{x})\mathcal{J}^{-1/2} \\ \nonumber &=& \hat{\mathbf{\Pi}}^a (\mathbf{x}) + \frac{i}{2}\frac{\delta \ln \mathcal{J}}{\delta \hat{\mathbf{A}}^a (\mathbf{x})}
\end{eqnarray}
we find that the quantum Hamiltonian density features the Faddeev-Popov determinant
\begin{eqnarray}
\label{shethami}
\hat{\mathscr{H}}_{\mathrm{YM}}&=&\frac{1}{2}\mathcal{J}^{-1}\hat{\mathbf{\Pi}}^a(\textbf{x})\mathcal{J}\hat{\mathbf{\Pi}}^a(\textbf{x})+\frac{1}{2}\hat{\textbf{B}}^a(\textbf{x})\cdot\hat{\textbf{B}}^a(\textbf{x})\\&+&\frac{1}{2}
\mathrm{g}^2\mathcal{J}^{-1}\hat{\varrho}_\mathfrak{g}^{b}\mathcal{J}\hat{\mathrm{M}}^{ba}(-\nabla^2)\hat{\mathrm{M}}^{ac}\hat{\varrho}_\mathfrak{g}^c.\nonumber
\end{eqnarray} 
Now the dynamical operators  in Eq.~(\ref{shethami}) are $\hat{\mathbf{A}}^a(\textbf{x})$
and  $\hat{\mathbf{\Pi}}^a(\textbf{x})=-i\delta/\delta\hat{\mathbf{A}}^a(\textbf{x})$, that do satisfy the  equal-time commutation relation
\begin{equation}\label{canocommrel}
\left[\hat{\mathrm{A}}_i^a(\textbf{x}),\hat{\Pi}_j^b(\textbf{y})\right]=i\delta^{ab}\left(\delta_{ij}-\frac{\nabla_i\nabla_j}{\nabla^2}\right)\delta^3(\textbf{x}-\textbf{y}).
\end{equation} 
A prize to pay is that $\hat{\mathbf{\Pi}}^a(\mathbf{x})$ is not a Hermitian operator for the curvilinear scalar product. Indeed,  $\hat{\mathbf{\Pi}}^{a\dagger}(\mathbf{x})=\mathcal{J}^{-1}\hat{\mathbf{\Pi}}^a(\mathbf{x})\mathcal{J}\neq \hat{\mathbf{\Pi}}^a(\mathbf{x})$ (see details in appendix \ref{fadeevandhermeticity}.)

We turn our attention to  $\boldsymbol{\mathcal{P}}_{\mathrm{YM}}$
defined in Eq. (\ref{gjdoa}). Taking into account Eq.
(\ref{gaugesdlsdiif}) we can express
\begin{equation}
\boldsymbol{\mathcal{P}}_{\mathrm{YM}}= \int \mathrm{d}^3
\mathrm{x}\mathrm{E}^{\ell a}(\textbf{x},\mathrm{t})\boldsymbol{\nabla} \mathrm{A}^{\ell a}(\textbf{x},\mathrm{t})
\end{equation}
The structure of the above expression does not  differ from those
obtained for the case of a pure $\rm U(1)-$gauge theory. New here is
the  sum running over the color indices.

We then split the chromoelectric field into transverse and
longitudinal part $\textbf{E}^a\equiv
\textbf{E}^a_{\mathrm{tr}}+\boldsymbol{\nabla}\Omega^a.$ From a
classical point of view
$\textbf{E}^a_{\mathrm{tr}}=-\boldsymbol{\pi}^a.$ Under these
conditions
\begin{eqnarray}\label{asdhoa}
\begin{array}{c}
\boldsymbol{\mathcal{P}}_{\mathrm{YM}}=-\int \mathrm{d}^3
\mathrm{x}\left\{\pi^{\ell a}(\textbf{x},\mathrm{t})\boldsymbol{\nabla}\mathrm{A}^{\ell a}(\textbf{x},\mathrm{t})\right.\\ \\ +\left.\partial^{\ell}\Omega^a(\textbf{x},\mathrm{t})\mathrm{A}^{\ell a}(\textbf{x},\mathrm{t})\right\}. 
\end{array}
\end{eqnarray} Next, we integrate by part the second
term of Eq.  (\ref{asdhoa}) and  use the identity
$\boldsymbol{\nabla}\cdot\mathbf{A}^a=0.$  As a consequence, it vanishes
identically and only the first term of Eq. (\ref{asdhoa}) remains. The quantized version in the  flat gauge field configuration  takes the form
\begin{equation}
\hat{\tilde{\boldsymbol{\mathcal{P}}}}_{\mathrm{YM}}=-\int \rm d^3
x\hat{\tilde{\Pi}}^{\ell a}(\textbf{x},t)\boldsymbol{\nabla}
\hat{\mathrm{A}}^{\ell a}(\textbf{x},t)  \label{asdhoa1}
\end{equation}
Finally, after applying the transformation 
\begin{equation} 
\hat{\boldsymbol{\mathcal{P}}}_{\mathrm{YM}} = \mathcal{J}^{-1/2} \hat{\tilde{\boldsymbol{\mathcal{P}}}}_{\mathrm{YM}}\mathcal{J}^{1/2}
\end{equation} and by considering Eq. (\ref{moemntumtransformationseps}) we  obtain 
\begin{equation}
\hat{\boldsymbol{\mathcal{P}}}_{\mathrm{YM}}=-\int \rm d^3
x\hat{\Pi}^{\ell a}(\textbf{x},t)\boldsymbol{\nabla}
\hat{\mathrm{A}}^{\ell a}(\textbf{x},t)  \label{asdhoa1}
\end{equation} (in agreement with~\cite{Besting:1989nq} and
generalizing the momentum  vector of electrodynamics).

%%%%%%%%%%%%%%%%%%%%%%%%%%%%%%%%%%%%%%%%%%%%%%%%%%%%%%%%%%%%%%%%%%%%%%%%%%%%
\subsection{Rotation and Boost Generators of  pure Yang-Mills theory}\label{subsec:poincare}
%%%%%%%%%%%%%%%%%%%%%%%%%%%%%%%%%%%%%%%%%%%%%%%%%%%%%%%%%%%%%%%%%%%%%%%%%%%%

In this subsection we obtain the generators associated to the
homogeneous Lorentz group.  We again invoke Noether's current for
the field $\rm A^\lambda(x),$ (with special transformation $\delta
\rm A^\lambda=1/2\omega^{\alpha\beta}
\left(\mathscr{J}_{\alpha\beta}\right)^{\lambda}_{\sigma}\rm
A^{\sigma})$
\begin{eqnarray}
\mathcal{J}_{\mathrm{YM}}^{\mu\alpha \beta}(\rm x) &=&\frac{\partial
\mathscr{L}_{\mathrm{YM}}}{\partial (\partial_{\mu}\rm A^{\nu a})}
\left\{\partial_{\lambda}\rm A^{\nu a} (\mathscr{J}^{\alpha
\beta})^\lambda_\sigma x^\sigma\right. \\&-&\left.
(\mathscr{J}^{\alpha\beta})^\nu_{\varsigma} \rm
A^{\varsigma a}\right\}-(\mathscr{J}^{\alpha\beta})_\nu^\mu x^\nu
\mathscr{L}_{\mathrm{YM}} \nonumber
\end{eqnarray} where $(\mathscr{J}_{\alpha\beta})^\ell_{\bar{\ell}}=\delta^\ell_\alpha
g_{\beta\bar{\ell}} -\delta^\ell_\beta g_{\alpha\bar{\ell}}$ is
the vectorial representation of the Lie algebra generator of $\rm SO(3,1).$
Recalling Eq.~(\ref{gaugesdlsdiif}) we can write
\begin{eqnarray}
\begin{array}{c}
\mathcal{J}_{\mathrm{YM}}^{\mu\alpha \beta}=\mathfrak{L}^{\mu\alpha\beta}+\mathfrak{S}^{\mu\alpha\beta},\\ \\ 
\mathfrak{L}^{\mu\alpha\beta}_{\mathrm{YM}}=\rm x^\alpha\mathscr{T}_{\mathrm{YM}}^{\mu\beta}-x^\beta\mathscr{T}_{\mathrm{YM}}^{\mu\alpha},\\ \\ \mathfrak{S}^{\mu\alpha\beta}=\mathfrak{F}^{\mu\beta a}\mathrm{A}^{\alpha a}-\mathfrak{F}^{\mu\alpha a}\mathrm{A}^{\beta a}.
\end{array}\label{curentss}
\end{eqnarray} Here  $\mathfrak{L}^{\lambda\mu\nu}$  is 
the orbital part whereas  $\mathfrak{S}^{\lambda\mu\nu}$ corresponds to the intrinsic spin. Note that $\mathcal{J}_{\mathrm{YM}}^{\mu\alpha \beta}$ include both rotations and boosts belonging to the Lorentz group. We will start by extracting the first. This is  obtained from Eq.~(\ref{curentss}) by fixing   $\mu=0$ and taking the spatial part of the remaining tensor. As a consequence the structure of the angular momentum  does not differ from an Abelian group except by a sum running over the color indices
\begin{eqnarray}
\begin{array}{c}\displaystyle
\boldsymbol{\mathfrak{J}}_\mathrm{YM}=\boldsymbol{\mathcal{L}}_{\mathrm{YM}}+\boldsymbol{\mathcal{S}}_{\mathrm{YM}},\\ \\ \displaystyle \boldsymbol{\mathcal{L}}_{\mathrm{YM}}=\int \rm d^3x \left\{ \mathbf{x}\times
\left[E^{\ell a}(\mathbf{x},t)\boldsymbol{\nabla}
\mathrm{A}^{\ell a}(\mathbf{x},t)\right]\right\},\\ \\ \displaystyle\boldsymbol{\mathcal{S}}_{\mathrm{YM}}=-\int \rm d^3x \left\{ \mathbf{E}^a(\mathbf{x},t)\times
\mathbf{A}^a(\mathbf{x},t)\right\}.
\end{array}
\end{eqnarray}
where we have used  $\mathfrak{F}^{i0a}=\mathrm{E}^{ai}.$
We again decompose $\mathbf{E}^a\to-\boldsymbol{\pi}^a+\boldsymbol{\nabla}\Omega^a.$ Terms arising from the longitudinal part,  $\boldsymbol{\nabla}\Omega^a$ cancel each other.  In the end,
\begin{eqnarray}\label{ROt1}
\begin{array}{c}\displaystyle
\hat{\boldsymbol{\mathfrak{J}}}_\mathrm{YM}=-\int \rm d^3x \left\{ \mathbf{x}\times
\left[\hat{\Pi}^{\ell a}(\mathbf{x},t)\boldsymbol{\nabla}\hat{\mathrm{A}}^{\ell a}(\mathbf{x},t)\right]\right.\\ \\ -\left.\hat{\boldsymbol{\Pi}}^a(\mathbf{x},t)\times
\hat{\mathbf{A}}^a(\mathbf{x},t)\right\}.
\end{array}
\end{eqnarray}

Next we address  the Boost generator in pure Yang-Mills theory,
fixing  $\mu=\alpha=0$ in Eq. (\ref{curentss}). This procedure leads
to
\begin{equation}
\mathcal{J}_\mathrm{YM}^{00i} = \rm x^0\mathscr{T}_\mathrm{YM}^{0i} -
x^i\mathscr{T}_\mathrm{YM}^{00}+\mathfrak{F}^{i0a}\mathrm{A}_0^a \ ,
\label{quququququq}
\end{equation} and
$\mathscr{T}_{\mathrm{YM}}^{00}$ must be understood as the
Hamiltonian density in Eq. (\ref{YMHamiltonia12n1d}). Quantizing
again at $\rm t=0$,  the corresponding charge is
\begin{eqnarray}
\begin{array}{c}\displaystyle
\boldsymbol{K}_{\mathrm{YM}}=\boldsymbol{\mathcal{K}}_{\mathrm{YM}}+\boldsymbol{\mathfrak{K}}_{\mathrm{YM}}, \\ \\ \displaystyle
\boldsymbol{\mathcal{K}}_{\mathrm{YM}}= - \int \mathrm{d}^3\mathrm{x} \left\{\mathbf{x}\left(
\frac{1}{2}\mathbf{B}^a\cdot\mathbf{B}^a +
\frac{1}{2}\boldsymbol{\pi}^{a}\cdot \boldsymbol{\pi}^a\right.\right. \\ + \displaystyle \left.\left. \frac{1}{2} \mathrm{g}^2
\varrho_\mathfrak{g}^{b}\mathrm{M}^{ba}(-\nabla^2)\mathrm{M}^{ac}\varrho_\mathfrak{g}^c\right)\right\}\\ \\
\boldsymbol{\mathfrak{K}}_{\mathrm{YM}}= \int\mathrm{d}^3\mathrm{x}\mathbf{E}^a(\mathbf{x})\mathrm{A}_0^a(\mathbf{x}).
\end{array}\label{Boost_spinpiece}
\end{eqnarray}
Returning to Eq.~(\ref{LagrangianYM}), the classical Euler-Lagrange
equation of motion associated to $\rm A_0$, Gauss's law, is a
constraint imposed on the quantization (above in
Eq.~(\ref{FPappears}) we used it in the form of the classical  Poisson
equation)
\begin{equation}
\boldsymbol{\nabla }\cdot \textbf{E}^a-\mathrm{gf}^{
abc}\textbf{A}^b\cdot \textbf{E}^c
=\mathbf{D}^{ab}\cdot\mathbf{E}^b=0 \label{gausslawPRI}
\end{equation}
Substitution of $\textbf{E}^a=-\dot{\textbf{A}}-\textbf{D}^{ab}\rm
A_0^b$ in the first term of Eq. (\ref{gausslawPRI}) returns
\begin{equation}
-\boldsymbol{\nabla}\cdot \textbf{D}^{\rm ab}\rm A_0^b =
g\varrho_\mathfrak{g}^a \label{e444}
\end{equation}
where we have used  the  trasversality condition $(\boldsymbol{\nabla}\cdot
\textbf{A}^a=0)$  and identified
$\varrho_\mathfrak{g}^a=\mathrm{f}^{abc}\mathbf{A}^b\cdot\boldsymbol{\pi}^c$.
Once more we  decompose the electric field into transverse and
longitudinal parts
\begin{equation}
\mathbf{E}^a=-\boldsymbol{\pi}^a+\boldsymbol{\nabla}\Omega^{a}\label{decopmpo}
\end{equation}  in the middle term  of Eq. (\ref{gausslawPRI}) to derive  the equation
\begin{equation}
\mathbf{D}^{ab}\cdot\boldsymbol{\nabla}\Omega^b=\mathrm{g}\varrho_g^a\label{sdfr456}
\end{equation}
(at this point $\rm A_0$ and $\Omega$ are interchangeable, but we
keep the distinction for another instant). Taking into account  Eq.
(\ref{decopmpo})  the spin piece of the boost,
Eq.~(\ref{Boost_spinpiece}), becomes 
\begin{eqnarray}
\begin{array}{c}
\boldsymbol{\mathfrak{K}}_{\mathrm{YM}}=\boldsymbol{\mathfrak{K}}_{\perp}+\boldsymbol{\mathfrak{K}}_{\parallel},\\  \\ \displaystyle
\boldsymbol{\mathfrak{K}}_{\perp}=-\mathrm{g}\int \mathrm{d}^3\mathrm{x}
\boldsymbol{\pi}^{a}\mathrm{M}^{ab}\varrho_\mathfrak{g}^a,\\ \\\displaystyle \boldsymbol{\mathfrak{K}}_{\parallel}=\mathrm{g}\int \mathrm{d}^3\mathrm{x} (\boldsymbol{\nabla}\Omega^{a})\mathrm{A}_0^a.
\end{array}
\label{falfhjsdfhj}
\end{eqnarray}
We express  $\boldsymbol{\mathfrak{K}}_{\perp}$ in  a more symmetric form
\begin{eqnarray}
\rm \boldsymbol{\mathfrak{K}}_{\perp}&=&-\frac{\mathrm{g}}{2}\int \mathrm{d}^3\mathrm{x}
\boldsymbol{\pi}^{a}\mathrm{M}^{ab}\varrho_\mathfrak{g}^b-\frac{\mathrm{g}}{2}\int \mathrm{d}^3\mathrm{x}\varrho_\mathfrak{g}^{b}
\mathrm{M}^{ba}\boldsymbol{\pi}^a. \label{falfhjsdfhj}
\end{eqnarray}

As for the longitudinal part, Eq.~(\ref{falfhjsdfhj}), substitution
of solutions of Eq.(\ref{sdfr456}) and Eq. (\ref{e444}) give
\begin{eqnarray}
\boldsymbol{\mathfrak{K}}_{\parallel}=-\mathrm{g}^2\int \mathrm{d}^3\mathrm{x}\boldsymbol{\nabla}(
\varrho_\mathfrak{g}^{b}\mathrm{M}^{ba})\mathrm{M}^{ac}\varrho^c_\mathfrak{g}\
.\label{sdasdxzsd}
\end{eqnarray}
After integration by parts   $\boldsymbol{\mathfrak{K}}_{\parallel}=\mathrm{g}^2\int \mathrm{d}^3\mathrm{x} \varrho_\mathfrak{g}^{b}\mathrm{M}^{ba}\boldsymbol{\nabla}(\mathrm{M}^{ac}\varrho^c_\mathfrak{g}).$ % On the other hand $\textbf{K}_s^{(l)}$ must be Hermitian which leads to  $\textbf{K}_s^{(l)\dagger}=\textbf{K}_s^{(l)}=g^2\int d^3x \varrho_\mathfrak{g}^{b}\mathrm{M}^{ba}\boldsymbol{\nabla}(\mathrm{M}^{ac}\varrho^c_\mathfrak{g})$.
Therefore $\rm \boldsymbol{\mathfrak{K}}_{\parallel}$ is a total derivative and
vanishes identically.

Wrapping up, the combination of Eq. (\ref{Boost_spinpiece}) and  Eq.
(\ref{falfhjsdfhj}) in $\mathbf{K}_{\mathrm{YM}}$  allows us to
express it as
\begin{eqnarray}\label{boostclassic}
\mathbf{K}_{\mathrm{YM}}=&-& \int \mathrm{d}^3\mathrm{x}
\left\{\mathbf{x}\left(\frac{1}{2}\mathbf{B}^a\cdot\mathbf{B}^a
+\frac{1}{2}\boldsymbol{\pi}^{a}\cdot \boldsymbol{\pi}^a\right.\right.\\&+&\left.\frac{\mathrm{g}^2}{2}
\varrho_\mathfrak{g}^{b}\mathrm{M}^{ba}(-\nabla^2)\mathrm{M}^{ac}\varrho_\mathfrak{g}^c\right)+\frac{\rm
g}{2}\boldsymbol{\pi}^{a}\mathrm{M}^{ab}\varrho_\mathfrak{g}^b\nonumber\\&+&\left.\frac{\rm
g}{2}\varrho_\mathfrak{g}^{b}
\mathrm{M}^{ba}\boldsymbol{\pi}^a\right\}.\nonumber
\end{eqnarray}

To match the expression of Besting and Sch\"utte it suffices to add
to Eq. (\ref{boostclassic})  a vanishing term proportional to
$\boldsymbol{\mathfrak{K}}_\parallel$: $0=\mathcal{N}\mathrm{g}^2\int \mathrm{d}^3\mathrm{x}\boldsymbol{\nabla}(
\varrho_\mathfrak{g}^{b}\mathrm{M}^{ba})\mathrm{M}^{ac}\varrho^c_\mathfrak{g}$,
and  choosing $\mathcal{N}=1/2$,   so that the corresponding  boost operator for a flat scalar product  can be written as
\begin{eqnarray}\label{boostclassic1}
\hat{\tilde{\mathbf{K}}}_{\mathrm{YM}}=&-& \int\mathrm{d}^3\mathrm{x}
\left\{\mathbf{x}\left(\frac{1}{2}\hat{\mathbf{B}}^a\cdot\hat{\mathbf{B}}^a
+\frac{1}{2}\hat{\tilde{\mathbf{\Pi}}}^{a\dagger}\cdot
\hat{\tilde{\mathbf{\Pi}}}^{a}\right)\right. - \\ \nonumber  &-&\frac{\mathrm{g}^2}{2}
\hat{\tilde{\varrho}}_\mathfrak{g}^{b\dagger}\hat{\mathrm{M}}^{ba}(\partial_\ell\mathbf{x}\partial_\ell)\hat{\mathrm{M}}^{ac}\hat{\tilde{\varrho}}_\mathfrak{g}^c+\frac{\mathrm{g}}{2}\hat{\tilde{\mathbf{\Pi}}}^{a\dagger}\hat{\mathrm{M}}^{ab}\hat{\tilde{\varrho}}_\mathfrak{g}^b +\nonumber\\&+&\left.\rm
\frac{g}{2}\hat{\tilde{\varrho}}_\mathfrak{g}^{b\dagger}\hat{\mathrm{M}}^{ba}\hat{\tilde{\mathbf{\Pi}}}^a\right\}.
\end{eqnarray}   

Making the change $
\hat{\mathbf{K}}_{\mathrm{YM}}=\mathcal{J}^{-1/2}\hat{\tilde{\mathbf{K}}}_{\mathrm{YM}}\mathcal{J}^{1/2}$ and minding the correct ordering of the various, non--commuting operators we obtain
\begin{eqnarray}\label{Boost1}
\begin{array}{c}\displaystyle
\hat{\mathbf{K}}_{\mathrm{YM}} =-\int \rm d^3
x\left\{\frac{1}{2}\mathcal{J}^{-1} \hat{\mathbf{\Pi}}^a \mathcal{J}
\mathbf{x}\hat{\mathbf{\Pi}}^a+\frac{1}{2}\hat{\mathbf{B}}^a
\mathbf{x} \hat{\mathbf{B}}^a\right.\\ \\
-\frac{1}{2}\mathrm{g}^2\mathcal{J}^{-1}\hat{\varrho}_\mathfrak{g}^{b}\mathcal{J}
\hat{\mathrm{M}}^{ba}(\partial_k \mathbf{x}\partial_k)
\hat{\mathrm{M}}^{ac}\hat{\varrho}_\mathfrak{g}^c\\ \\
+\left. \rm \frac{1}{2}g\mathcal{J}^{-1}\hat{\mathbf{\Pi}}^a
\mathcal{J}\hat{\mathrm{M}}^{ab}\hat{\varrho}_\mathfrak{g}^b+\frac{1}{2}g
\mathcal{J}^{-1}\hat{\varrho}_\mathfrak{g}^{b}\mathcal{J}
\hat{\mathrm{M}}^{ba}\hat{\mathbf{\Pi}}^a\right\}.
\end{array}
\end{eqnarray}
Eq. (\ref{Boost1})  coincides with the expression derived in  \cite{Besting:1989nq}. To derive the above expression we have replaced the operator $\hat{\tilde{\mathbf{\Pi}}}^a(\textbf{x})$ by the transformed one (see Eq. (\ref{moemntumtransformationseps})).

%%%%%%%%%%%%%%%%%%%%%%%%%%%%%%%%%%%%%%%%%%%%%%%%%%%%%%%%%%%%%%%%%%%%%%%%%%%%%%%%%%%
\subsection{Inclusion of the Quark sector}\label{subsec:quark}
%%%%%%%%%%%%%%%%%%%%%%%%%%%%%%%%%%%%%%%%%%%%%%%%%%%%%%%%%%%%%%%%%%%%%%%%%%%%%%%%%%%

To complete the discussion on Quantum Chromodynamics we include  the  matter  sector interacting with the gauge field. This is achieved  by the matter Lagragian
\begin{equation}
\mathscr{L}_{\mathrm{Matter}}=\hat{\bar{\mathfrak{q}}}^\ell(i\gamma^\mu\partial_\mu-\mathrm{m})\hat{\mathfrak{q}}^\ell+\mathrm{g}\hat{\bar{\mathfrak{q}}}^\ell\gamma^\mu\frac{\lambda^a}{2}\hat{\mathrm{A}}_\mu^a\hat{\mathfrak{q}}^\ell
\end{equation} with $\hat{\bar{\mathfrak{q}}}=\hat{\mathfrak{q}}^\dagger\gamma^0$ and   $\lambda^a/2$  representing the $\rm SU(3)$ Gell-Mann matrices.

Its contribution to the stress-energy tensor is
\begin{eqnarray}\label{matterSET}
\mathscr{T}_{\mathrm{Matter}}^{\mu\nu}&=&\frac{\partial \mathscr{L}_{\mathrm{Matter}}}{\partial(\partial_\mu\hat{\mathfrak{q}}^\ell)}\partial^\nu\hat{\mathfrak{q}}^\ell-g^{\mu\nu}\mathscr{L}_{\mathrm{Matter}}\\&=&i\hat{\bar{\mathfrak{q}}}^\ell\gamma^\mu\partial^\nu\hat{\mathfrak{q}}^\ell-g^{\mu\nu}\hat{\bar{\mathfrak{q}}}^\ell(i\gamma^\lambda\partial_\lambda-\mathrm{m})\hat{\mathfrak{q}}^\ell- \\ \nonumber
&-&\mathrm{g}g^{\mu\nu}\hat{\bar{\mathfrak{q}}}^\ell\gamma^\mu\frac{\lambda^a}{2}\hat{\mathrm{A}}_\mu^a\hat{\mathfrak{q}}^\ell\nonumber
\end{eqnarray}The conserved charges associated to space-time translational invariance are then
\begin{eqnarray}
\begin{array}{c}
\hat{\mathrm{H}}_{\mathrm{Matter}}=\int\mathrm{d}^3 \mathrm{x} \mathscr{T}^{00}_{\mathrm{Matter}}=\int\mathrm{d}^3 \mathrm{x}\left\{ \hat{\mathscr{H}}_{\mathrm{Dirac}}+\hat{\mathscr{H}}_{\mathrm{Int}}\right\},\\ \\ 
\hat{\boldsymbol{\mathcal{P}}}_{\mathrm{Matter}}=\int \mathrm{d}^3 \mathrm{x}\mathscr{T}^{0i}_{\mathrm{Matter}}=\int \mathrm{d}^3 \mathrm{x} \hat{\mathfrak{q}}^{\ell\dagger } (\mathbf{x})\left( - i \boldsymbol{\nabla}\right) \hat{\mathfrak{q}}^\ell (\mathbf{x})
\end{array}\label{pqwsalmhjf}
\end{eqnarray}
where   $\hat{\mathscr{H}}_\mathrm{Dirac}$ is the free Dirac Hamiltonian density:
\begin{equation}
\hat{\mathscr{H}}_\mathrm{Dirac}(\mathbf{x})=\hat{\mathfrak{q}}^{\ell\dagger}(\mathbf{x}) (-i \boldsymbol{\alpha}\cdot \boldsymbol{\nabla} + \beta \mathrm{m} ) \hat{\mathfrak{q}}^\ell(\mathbf{x}),
\end{equation} whereas
\begin{eqnarray}\label{hinrer}
\hat{\mathscr{H}}_{\mathrm{Int}}(\mathbf{x})&=&\mathrm{g}\hat{\mathfrak{q}}^{\ell\dagger} (\mathbf{x}) \frac{\lambda^a}{2}\boldsymbol{\alpha}\cdot \hat{\mathbf{A}}^a(\mathbf{x}) \hat{\mathfrak{q}}^\ell(\mathbf{x})\\&-&\mathrm{g}\hat{\mathfrak{q}}^{\ell\dagger} (\mathbf{x}) \frac{\lambda^a}{2}\hat{\mathfrak{q}}^\ell(\mathbf{x}) \hat{\sigma}^a(\mathbf{x}) \nonumber
\end{eqnarray}
arises due to the interaction between the gauge field and Quarks.  The last term of the above equation introduces  the quark color charge density $\hat{\varrho}_\mathfrak{q}^a=\hat{\mathfrak{q}}^{\ell\dagger}\frac{\lambda^a}{2} \hat{\mathfrak{q}}^{\ell}$
that must be summed to the right hand side (source) of Poisson's equation in
Eq.~(\ref{omesoliuti}) and following. As a consequence $\Omega^a\to\hat{\Omega}^a=\hat{\mathrm{M}}^{ab}\hat{\varrho}^b$ with $\hat{\varrho}^a=\hat{\varrho}_\mathfrak{g}^a+\hat{\varrho}_\mathfrak{q}^a.$
The second term of Eq.~(\ref{hinrer}) adds up to the second term of Eq.~(\ref{shethami}) and both together have the same functional form as the latter
but with $\hat{\varrho}_\mathfrak{g}^a$ replaced $\hat{\varrho}^a=\hat{\varrho}_\mathfrak{g}^a+\hat{\varrho}_\mathfrak{q}^a$.

The complete QCD Hamiltonian and momentum  densities are the sum of both
\begin{eqnarray}
\hat{\rm H}_\mathrm{QCD} &=&\hat{\rm H}_\mathrm{YM}+\hat{\rm H}_\mathrm{Matter},\\
\hat{\boldsymbol{\mathcal{P}}}_{\mathrm{QCD}}&=&\hat{\boldsymbol{\mathcal{P}}}_{\mathrm{YM}}+\hat{\boldsymbol{\mathcal{P}}}_{\mathrm{Matter}}.
\label{fhtg}
\end{eqnarray} where  $\hat{\rm H}_\mathrm{YM}$ is the corresponding operational version of  Eq. (\ref{YMHamiltonia12n1d}) with  $\hat{\mathscr{H}}_\mathrm{YM}$  given by Eq. (\ref{shethami})
In turn, Eq. (\ref{fhtg}) combines Eq.~(\ref{asdhoa1}) and Eq.~(\ref{pqwsalmhjf}).

We now turn to the angular momentum and boost operators. The conserved current due to quarks is obtained from the tensor
\begin{eqnarray} \mathcal{J}_{\mathrm{Matter}}^{\mu\alpha\beta}&=&\mathrm{x}^\alpha\mathscr{T}^{\mu\beta}_{\mathrm{Matter}}-\mathrm{x}^\beta\mathscr{T}^{\mu\alpha}_{\mathrm{Matter}}+ \frac{i}{4}\frac{\partial
\mathscr{L}_{\mathrm{Matter}}}{\partial(\partial_\mu \hat{\mathfrak{q}}^{\ell})}\sigma^{\alpha\beta}\hat{\mathfrak{q}}^\ell\nonumber\\&-& \frac{i}{4}\hat{\bar{\mathfrak{q}}}^{\ell}\sigma^{\alpha\beta}\frac{\partial
\mathscr{L}_{\mathrm{Matter}}}{\partial(\partial_\mu \hat{\bar{\mathfrak{q}}}^{\ell})}.
\label{sdasd2}
\end{eqnarray}
Here $\partial \mathscr{L}_{\mathrm{Matter}}/\partial(\partial_\mu \hat{\mathfrak{q}}^{\ell})=i\hat{\bar{\mathfrak{q}}}\gamma^\mu$    whereas
$\partial \mathscr{L}_{\mathrm{Matter}}/\partial(\partial_\mu \hat{\bar{\mathfrak{q}}}^{\ell})=-i\gamma^\mu\hat{\mathfrak{q}}$ therefore
\begin{equation} 
\begin{array}{c}\displaystyle
\mathcal{J}_{\mathrm{Matter}}^{\mu\alpha\beta}(\mathrm{x})=\mathfrak{L}^{\mu\alpha\beta}_{\mathrm{Matter}}+\mathfrak{S}^{\mu\alpha\beta}_{\mathrm{Matter}},\\ \\\displaystyle \mathfrak{L}^{\mu\alpha\beta}_{\mathrm{Matter}}\equiv \mathrm{x}^\alpha\mathscr{T}^{\mu\beta}_{\mathrm{Matter}}-\mathrm{x}^\beta\mathscr{T}^{\mu\alpha}_{\mathrm{Matter}},\\ \\ \mathfrak{S}^{\mu\alpha\beta}_{\mathrm{Matter}}=-\frac{1}{4}\hat{\bar{\mathfrak{q}}}^{\ell}\left\{\gamma^\mu,\sigma^{\alpha\beta}\right\}\hat{\mathfrak{q}}^\ell.\end{array}
\label{sdasd3f}
\end{equation}

By fixing $\mu=0$ and taking the remaining indices as spatial, we obtain the quark angular momentum, $\mathfrak{J}_{\mathrm{Matter}}^i=1/2\epsilon^{ijk}\int \mathrm{d}^3\mathrm{x}(\mathrm{x}^j\mathscr{T}_{\mathrm{Matter}}^{0k}-\frac{1}{2}\hat{\mathfrak{q}}^{\ell\dagger}\sigma^{jk}\hat{\mathfrak{q}}^\ell)$, where $\mathfrak{T}_{\mathrm{Matter}}^{0k}$ is the momentum density in Eq. (\ref{pqwsalmhjf}).
Therefore\begin{eqnarray}\label{sdasd4f}
\begin{array}{c}
\hat{\boldsymbol{\mathfrak{J}}}_\mathrm{Matter}=\hat{\boldsymbol{\mathcal{L}}}_\mathrm{Matter}+\hat{\boldsymbol{\mathcal{S}}}_\mathrm{Matter},\\ \\ \displaystyle \hat{\boldsymbol{\mathcal{L}}}_\mathrm{Matter}=\int \mathrm{d}^3\mathrm{x}\left\{\hat{\mathfrak{q}}^{\ell\dagger} (\mathbf{x})\left[ \mathbf{x} \times (-i\boldsymbol{\nabla})\right] \hat{\mathfrak{q}}^\ell(\mathbf{x})\right\},\\ \\  \hat{\boldsymbol{\mathcal{S}}}_\mathrm{Matter}=\int \mathrm{d}^3\mathrm{x}\hat{\mathfrak{q}}^{\dagger \ell} (\mathbf{x}) \left( \frac{1}{2} \boldsymbol{\Sigma}
\right) \hat{\mathfrak{q}}^\ell(\mathbf{x}),
\end{array}
\end{eqnarray} with $\Sigma_i=\frac{1}{2}\epsilon_{ijk}\sigma_{jk}.$

Following a procedure similar to that used in deriving
Eq.~(\ref{quququququq}) one obtains the Boost density easily. The
generator is obtained integrating over the $(\rm t=0)$ surface
$\hat{K}_{\mathrm{Matter}}^{i}=\int
\mathrm{d}^3\mathrm{x}\left\{-\mathrm{x}^i\mathscr{T}_{\mathrm{Matter}}^{00}\right.$
$-\left.\frac{1}{2}\hat{\mathfrak{q}}^{\ell\dagger}\sigma^{0i}\hat{\mathfrak{q}}^\ell\right\}.$ Therefore
\begin{eqnarray}
\begin{array}{c}
\hat{\boldsymbol{K}}_{\mathrm{Matter}}=\hat{\boldsymbol{\mathcal{K}}}_{\mathrm{Matter}}+\hat{\boldsymbol{\mathfrak{K}}}_{\mathrm{Matter}},\\ \\  \displaystyle \hat{\boldsymbol{\mathcal{K}}}_{\mathrm{Matter}}=-\int \mathrm{d}^3\mathrm{x}\mathbf{x}\left\{\hat{\mathfrak{q}}^{\ell\dagger} (\mathbf{x}) \left( -i\boldsymbol{\alpha}\cdot \boldsymbol{\nabla} + \beta\mathrm{m} \right) \hat{\mathfrak{q}}^\ell(\mathbf{x})\right.\\ \\ +\left.\frac{\mathrm{g}}{2}\hat{\mathfrak{q}}^{\ell\dagger} (\mathbf{x}) \lambda^a\boldsymbol{\alpha}\cdot \hat{\mathbf{A}}^a(\mathbf{x}) \hat{\mathfrak{q}}^\ell(\mathbf{x})\right\},\\ \\
\hat{\boldsymbol{\mathfrak{K}}}_{\mathrm{Matter}}=\int \mathrm{d}^3 \mathrm{x}\hat{\mathfrak{q}}^{\ell\dagger}(\mathbf{x})\left(\frac{i}{2}\boldsymbol{\alpha} \right) \hat{\mathfrak{q}}^\ell (\mathbf{x})
\end{array}
\end{eqnarray} with $\alpha_i=i\sigma_{0i}.$
For complete  QCD, one needs to substitute again
$\hat{\varrho}_\mathfrak{g}^a$ by
$\hat{\varrho}^a=\hat{\varrho}_\mathfrak{g}^a+\hat{\varrho}_\mathfrak{q}^a$
in  Eq. (\ref{ROt1}), the boost generator for pure Yang-Mills theory.
Finally one finds  
\begin{eqnarray}
\hat{\boldsymbol{\mathfrak{J}}}_\mathrm{QCD}&=&\hat{\boldsymbol{\mathfrak{J}}}_\mathrm{YM}+\hat{\boldsymbol{\mathfrak{J}}}_\mathrm{Matter},\\
\hat{\boldsymbol{K}}_{\mathrm{QCD}}&=&\hat{\boldsymbol{K}}_{\mathrm{YM}}
+\hat{\boldsymbol{K}}_{\mathrm{Matter}}.
\end{eqnarray}
(keeping in mind that $\hat{\varrho}_\mathfrak{g}^a\to\hat{\varrho}^a=\hat{\varrho}_\mathfrak{g}^a+\hat{\varrho}_\mathfrak{q}^a$).

We also record the (possibly Bogoliubov rotated) normal mode expansions
of the dynamical fields
\begin{eqnarray} \label{normalmodes}
\begin{array}{c}\displaystyle
\hat{\mathbf{A}}^a(\mathbf{x}) = \int \frac{\dbar\mathrm{k}}{\sqrt{2\omega_{\mathbf{k}}}}\left[\mathbf{a}^a(\mathbf{k}) +\mathbf{a}^{a\dagger}(-\mathbf{k}) \right] e^{i\mathbf{k}\cdot \mathbf{x}},\\ \\ \displaystyle
\hat{\boldsymbol{\Pi}}^a(\mathbf{x}) = -i\int \dbar\mathrm{k}\sqrt{\frac{\omega_{\mathbf{k}}}{2}}\left[ \mathbf{a}^a(\mathbf{k}) -\mathbf{a}^{a\dagger}(-\mathbf{k}) \right] e^{i\mathbf{k}\cdot \mathbf{x}},\\ \\
\displaystyle
\hat{\mathfrak{q}}^\ell (\mathbf{x}) = \sum_\lambda \int \dbar\mathrm{k} e^{i\mathbf{k}\cdot \mathbf{x}}\left( \mathscr{B}_{\mathbf{k}\lambda}^\ell \mathrm{U}_{\mathbf{k}\lambda} + \mathscr{D}^{\ell\dagger}_{-\mathbf{k}\lambda}\mathrm{V}_{-\mathbf{k}\lambda}\right),  \\ \\ \displaystyle
\hat{\mathfrak{q}}^{\ell\dagger} (\mathbf{x})= \sum_\lambda \int \dbar\mathrm{k} e^{-i\mathbf{k}\cdot \mathbf{x}}\left( \mathscr{B}^{\ell\dagger}_{\mathbf{k}\lambda}\mathrm{U}^\dagger_{\mathbf{k}\lambda} + \mathscr{D}_{-\mathbf{k}\lambda}^\ell \mathrm{V}^\dagger_{-\mathbf{k}\lambda} \right)\end{array}
\end{eqnarray} with $\dbar\mathrm{k}\equiv  \mathrm{d}^3\mathrm{k}/(2\pi)^3$ and $\omega_{\mathbf{k}}=\vert\mathbf{k}\vert$.
In momentum space   Eq. (\ref{canocommrel}) reduces to
\begin{equation}
\left[\mathrm{a}_i^a(\mathbf{k}),\mathrm{a}_j^{b\dagger}(\mathbf{k}^\prime)\right]=(2\pi)^3\delta^{ab}\left(\delta_{ij}-\frac{\mathrm{k}_i\mathrm{k}_j}{\mathbf{k}^2}\right)\delta^3(\mathbf{k}-\mathbf{k}')
\end{equation}
Transversality translates into $\mathbf{k}\cdot\mathbf{a}^a(\mathbf{k})=\mathbf{k}\cdot\mathbf{a}^{a\dagger}(\mathbf{k}).$ The operators $\mathbf{a}^a(\mathbf{k})$ and $\mathbf{a}^{a\dagger}(\mathbf{k})$ include the gluon polarization  $\mathbf{a}^{a}\equiv\sum_{r=1,2} \boldsymbol{\epsilon}_{r}(\mathbf{k})a^a_{r}(\mathbf{k})$ and the usual creation $a_{r}^{a\dagger}(\mathbf{k})$ and annihilation $a^a_{r}(\mathbf{k})$ operators (the normalization is as usual $\epsilon_r^i \epsilon_s^i=\delta_{rs}$, $\epsilon_r^i \epsilon_r^j= \delta^{ij}-\mathrm{k}^i \mathrm{k}^j/\mathbf{k}^2$).

Similarly, quark creation and annihilation operators satisfy the anticommutation relations
\begin{equation}
\left\{\mathscr{B}_{\mathbf{k}\lambda},\mathscr{B}_{\mathbf{k}'\lambda'}^\dagger\right\}=\left\{\mathscr{D}_{\mathbf{k}\lambda},\mathscr{D}_{\mathbf{k}'\lambda'}^\dagger\right\}=(2\pi)^3\delta^3(\mathbf{k}-\mathbf{k}^{\prime})\delta_{\lambda\lambda'}
\end{equation}
(others zero).

The renormalization of the boost generators also deserves a comment.
It is clear that radiative corrections to the classical, conformal
theory are infinite and the quantum theory is only well defined
after a scale is chosen. Regularization and renormalization in
fixed-gauge Hamiltonian dynamics is arduous and is being pursued
independently with attention to the Slavnov-Taylor
identities~\cite{Watson:2008fb}. One should like to find a
Lorentz-invariant regulator that would leave the Poincar\'e algebra
intact (which could be established by looking into Schwinger's
condition). Whatever strategy one adopts to regularize the
Hamiltonian, for example a lattice regularization, one can similarly
apply it to the boost operator. A large part of the same, coming
from $\rm x^iH$, is automatically regulated and requires no additional
counterterms other than those in $\rm H$, as the additional power of
$\rm x^i$ lowers the mass-dimension. This leaves the spin terms like
$\boldsymbol{\mathfrak{K}}$, and in principle, one should expect mass
counterterms of equal or lower mass dimension than those spin terms
to appear. The issue is postponed to future work.

%%%%%%%%%%%%%%%%%%%%%%%%%%%%%%%%%%%%%%%%%%%%%%%%%%%%%%%%%%%%%%%%%%%%%%%%%%%%%%%%%%%%%%%%%
\section{$\phi$--radiative decays} \label{sec:phidec}
%%%%%%%%%%%%%%%%%%%%%%%%%%%%%%%%%%%%%%%%%%%%%%%%%%%%%%%%%%%%%%%%%%%%%%%%%%%%%%%%%%%%%%%%%

%%%%%%%%%%%%%%%%%%%%%%%%%%%%%%%%%%%%%%%%%%%%%%%%%%%%%%%%%%%%%%%%%%%%%%%%%%%%%%%%%%%%%%%%%
\subsection{The boosted wavefunctions of decay products}
%%%%%%%%%%%%%%%%%%%%%%%%%%%%%%%%%%%%%%%%%%%%%%%%%%%%%%%%%%%%%%%%%%%%%%%%%%%%%%%%%%%%%%%%%

We consider again the radiative decay $\phi\to \gamma \eta$ as an example, but it is obvious that the discussion is general. The boosted meson with velocity $v$ is given (in terms of the rapidity $\zeta$ defined in Eq.~(\ref{rapidity}))  by
\begin{eqnarray}
\vert \eta_v \rangle &=&e^{i \hat{\mathbf{K}}\cdot\boldsymbol{\zeta}} \vert \eta_0 \rangle  \\ \nonumber
&\simeq& \vert \eta_0 \rangle + i \hat{\mathbf{K}}\cdot\boldsymbol{\zeta}  \vert \eta_0 \rangle
- \frac{(\hat{\mathbf{K}}\cdot\boldsymbol{\zeta})^2}{2}\vert \eta_0 \rangle \dots
 \end{eqnarray}
We want to establish that a gluonium term, not taken into account in past analysis, arises just because of the change of reference frame, even if $\vert \eta_0 \rangle$ contained only purely $\mathfrak{q}\bar{\mathfrak{q}}$ configurations. Therefore we need to see what pieces of the QCD boost operator connect the $\mathfrak{q}\bar{\mathfrak{q}}$ and $\mathfrak{g}\mathfrak{g}$ Fock subspaces.

At linear order in $v$, the two spaces are indeed disconnected, since the interacting part of $\hat{\mathbf{K}}$ that could effect the change is proportional to the color charge densities in quarks and gluons respectively, $\hat{\varrho}_\mathfrak{q} \hat{\varrho}_\mathfrak{g} \propto 1/2\hat{\mathfrak{q}}^\dagger\lambda^a \hat{\mathfrak{q}}\mathrm{f}^{abc} \hat{\textbf{A}}^b\cdot \hat{\boldsymbol{\Pi}}^c $, and upon closing the quark line, color arithmetic sets the contribution to zero since $\mathrm{Tr}(\lambda)=0$. One needs then to resource to the second order term, $\hat{\mathbf{K}}^2$, where the square of the instantaneous Coulomb potential contributes, as does the square of the gauge coupling to transverse gluons.

We focuse on the latter
\begin{eqnarray}
\label{boost_used}
\hat{\mathrm{K}}_z=- \frac{\mathrm{g}}{2} \int \mathrm{d}^3\mathrm{x} z \hat{\mathfrak{q}}^{\ell\dagger}(\textbf{x})\lambda^a \boldsymbol{\alpha}\cdot\hat{\textbf{A}}^a (\textbf{x})\hat{\mathfrak{q}}^\ell(\textbf{x})
\end{eqnarray}
to show the phenomenon, not because the contribution of the $\boldsymbol{\alpha}\cdot\hat{\textbf{A}}$ Hamiltonian is larger than the $\hat{\varrho}\hat{\varrho}$ piece, but just because of simplicity: this contribution is the one with the smallest number of loops that turns out not to vanish, which we here demonstrate. An added bonus is that, although formally in Coulomb gauge, the piece we calculate will be present in other gauges since it involves only the transverse gluons. The calculational methods are analogous to those of the estimate of the glueball width in Coulomb-gauge models of QCD~\cite{Bicudo:2006sd}.

With $v\simeq 0.55$ for our particular case, $\frac{\zeta^2}{2}\simeq 0.15$. For the rest--frame state is sufficient to take the color singlet, flavor singlet (through mixing) component of the $\eta$ meson
\begin{eqnarray}
\vert \eta_0 \rangle  &=&  \sin \left(\phi_\mathrm{P}\right)
\int\dbar^3\mathrm{k}\ \ \mathrm{Y}^0_0(\Omega_\mathbf{k})  \mathscr{F}_{\eta_0}(\vert \textbf{k}\vert)
\\ \nonumber
&\times&\sum_{\mu_1 \mu_2} \left\langle\frac{1}{2} \mu_1 \left.\frac{1}{2} -\mu_2\right\vert 0 0 \right\rangle
(-1)^{1/2+\mu_2}\ \  \mathscr{B}_{\mathbf{k}\mu_1}^\dagger \mathscr{D}_{-\mathbf{k}\mu_2}^\dagger \vert 0 \rangle
\end{eqnarray}with color wavefunction $\delta^{i_1 i_2} /(\rm N_c)^{1/2}$, and flavor wavefunction understood to be the singlet projection.

Substituting the normal--mode expansion Eq.~(\ref{normalmodes})  for the fields $\hat{\mathfrak{q}}$ and $\hat{\textbf{A}}^a$  in  Eq.~(\ref{boost_used}), one obtains, in terms of BCS spinors $\rm U$, $\rm V$,
a gluonium component of the $\eta$ moving with velocity $v$,
\begin{eqnarray}
\label{primerestado}
\vert \eta_v^\mathfrak{g} \rangle &=&  \mathrm{g}^2 \frac{\zeta^2}{2} \sin\left(\phi_\mathrm{P}\right)\! \! \int\ \ \!\dbar \mathrm{q}
\int\ \ \! \dbar \mathrm{p} \int\! \mathrm{d}^3\mathrm{k} \int\! \mathrm{d}^3 \mathrm{k}' \\ \nonumber
&\times&\partial_{\mathrm{p}_z} \left(\delta^{(3)}(\mathbf{k}+\mathbf{p}-\mathbf{q})\right)
\partial_{\mathrm{p}_z} \left(\delta^{(3)}(\mathbf{k}'+\mathbf{q}-\mathbf{p})\right)
\\ \nonumber
&\times&\sum_{\lambda \mu_1 \mu_2} \frac{1}{\sqrt{4\pi}} \frac{1}{2\sqrt{\omega_\mathbf{k} \omega_ {\mathbf{k}'}}} \left\langle\frac{1}{2} \mu_1 \left.\frac{1}{2} -\mu_2\right\vert 0 0 \right\rangle
(-1)^{1/2+\mu_2}
\\ \nonumber
&\times& \mathscr{F}_{\eta_0}(\vert\mathbf{q}\vert) \mathrm{V}^\dagger_{-\mathbf{q}\lambda} \alpha^j \frac{\lambda^a}{2} \mathrm{U}_{\mathbf{p}\mu_1} \mathrm{V}^\dagger_{-\mathbf{p}\mu_2}
\alpha^{j'} \frac{\lambda^b}{2} \mathrm{V}_{-\mathbf{q} \lambda}
a^{\dagger a j}_\mathbf{k} a^{\dagger b j'}_{\mathbf{k}'} \vert 0 \rangle \ .
\end{eqnarray}
The color algebra is easy since $\mathrm{Tr}(\lambda^a\lambda^b)= 2\delta^{ab}$.
The spin work however is a little more tedious. Let us adopt a definition of the spinors in terms of a BCS angle, defined as $s_\mathbf{k} \equiv \sin \varphi(\vert\mathbf{k}\vert) = \frac{\mathrm{m}(\vert\mathbf{k}\vert)}{\mathrm{E}(\vert\mathbf{k}\vert)}$, with $\mathrm{E}(\vert\mathbf{k}\vert)=[\mathbf{k}^2+\mathrm{m}(\vert\mathbf{k}\vert)^2]^{1/2}$ and $c_\mathbf{k}\equiv \cos\varphi(\vert\mathbf{k}\vert) = \frac{\vert\mathbf{k}\vert}{\mathrm{E}(\vert\mathbf{k}\vert)}$.
The spinors read then
\begin{eqnarray}
\mathrm{U}_{\mathbf{k}\lambda} &=& \frac{1}{2^{1/2}}
\left[ \begin{array}{c}
(1+s_\mathbf{k})^{1/2} \chi_\lambda \\
(1-s_\mathbf{k})^{1/2} \boldsymbol{\sigma}\cdot \mathbf{n}_\mathbf{k}\chi_\lambda
\end{array} \right] \\ \nonumber \\
\mathrm{V}_{-\mathbf{k}\lambda} &=& \frac{1}{2^{1/2}}
\left[ \begin{array}{c}-(1-s_\mathbf{k})^{1/2} \boldsymbol{\sigma}\cdot \mathbf{n}_{\mathbf{k}}\chi_\lambda \\
(1+s_\mathbf{k})^{1/2} \chi_\lambda
\end{array} \right]
\end{eqnarray}in terms of (bidimensional) Pauli matrices and Pauli spinors satisfying the closure relation $\sum_\lambda \chi_\lambda \chi_\lambda^\dagger = 1_{2\times 2}$. Here $\mathbf{n}_{\mathbf{k}}=\mathbf{k}/\vert\mathbf{k}\vert$ is 
a unit vector.

The  spin combination needed yields
\begin{equation}
\begin{split}
& 4\sum_\lambda \mathrm{V}^\dagger_{-\mathbf{q}\lambda}\alpha^j \mathrm{U}_{\mathbf{p}\mu_1}\mathrm{V}^\dagger_{-\mathbf{p}\mu_2}\alpha^{j'}\mathrm{V}_{-\mathbf{q}\lambda} = \\
&= \sum_\lambda \chi^\dagger_\lambda \left\{ (1-s_\mathbf{p})c_\mathbf{q} \boldsymbol{\sigma}\cdot\mathbf{n}_{\mathbf{q}}   \sigma^j \boldsymbol{\sigma}\cdot\mathbf{n}_{\mathbf{p}}\chi_{\mu_1}\chi^\dagger_{\mu_2}\boldsymbol{\sigma}\cdot \mathbf{n}_{\mathbf{p}}\sigma^{j'} \right.+ \\
&+(1-s_\mathbf{q})c_\mathbf{p} \boldsymbol{\sigma}\cdot\mathbf{n}_{\mathbf{q}}\sigma^j \boldsymbol{\sigma}\cdot\mathbf{n}_{\mathbf{p}}\chi_{\mu_1}\chi^\dagger_{\mu_2}\sigma^{j'}\boldsymbol{\sigma}\cdot \mathbf{n}_{\mathbf{q}} - \\
&  - (1+s_\mathbf{q})c_\mathbf{p}\sigma^j\chi_{\mu_1}\chi^\dagger_{\mu_2}\boldsymbol{\sigma}\cdot\mathbf{n}_{\mathbf{p}}\sigma^{j'} - \\
&-\left.(1+s_\mathbf{p})c_\mathbf{q} \sigma^j \chi_{\mu_1}\chi^\dagger_{\mu_2}\sigma^{j'}\mathbf{\sigma}\cdot \mathbf{n}_{\mathbf{q}} \right\} \chi_{\lambda}
\end{split}
\end{equation}

Then we take into account that the Clebsch-Gordan coefficient in Eq. (\ref{primerestado}) is
\begin{equation}
\left\langle\frac{1}{2} \mu_1 \left.\frac{1}{2} -\mu_2\right\vert 0 0 \right\rangle (-1)^{1/2+\mu_2} =
-\frac{1}{\sqrt{2}} \delta_{\mu_1 \mu_2}
\end{equation}
which forces a trace over the Pauli matrices.
Employing cyclicity of the trace and $\mathrm{Tr}(\sigma^i \sigma^j \sigma^{k})=
2i\epsilon^{ijk}$, and returning to the running mass $\mathrm{m}(\vert\mathbf{p}\vert)$ and energy $\mathrm{E}(\vert\mathbf{p}\vert)$, we have
\begin{eqnarray}
\ar \eta^\mathfrak{g}_v \ra&=& \frac{v^2}{2} (4\pi \alpha_s) \sin\left(\phi_\mathrm{P}\right)\! \! \int\! \dbar \mathrm{q} 
\int\! \dbar \mathrm{p} \\ \nonumber&\times&
\frac{1}{\sqrt{4\pi}} \frac{i\epsilon^{ijj'}}{\sqrt{2}} \frac{1}{2\sqrt{N_c}}
\frac{\mathrm{q}^i \mathrm{m}(\vert \mathbf{p}\vert)-\mathrm{p}^i \mathrm{m}(\vert\mathbf{q}\vert)}{\mathrm{E}(\vert\mathbf{q}\vert)\mathrm{E}(\vert\mathbf{p}\vert)} \mathscr{F}_{\eta_0}(\vert\mathbf{q}\vert)\\ \nonumber&\times&
\partial_{\mathrm{q}_z} \left( \frac{1}{\sqrt{2\omega_{\mathbf{q}-\mathbf{p}}}} a^{\da aj}_{\mathbf{q}-\mathbf{p}}  \right)
(-\partial_{\mathrm{q}_z}) \left( \frac{1}{\sqrt{2\omega_{\mathbf{p}-\mathbf{q}}}} a^{\da aj}_{\mathbf{p}-\mathbf{q}}  \right)\ar 0 \ra. 
\end{eqnarray}where $\alpha_s=\mathrm{g}^2/(4\pi)$ is the strong fine structure constant. 

%%%%%%%%%%%%%%%%%%%%%%%%%%%%%%%%%%%%%%%%%%%%%%%%%%%%%%%%%%%%%%%%%%%%%%%%%%%%%%%%%%%%%%%
\subsection{The gluonium content}
%%%%%%%%%%%%%%%%%%%%%%%%%%%%%%%%%%%%%%%%%%%%%%%%%%%%%%%%%%%%%%%%%%%%%%%%%%%%%%%%%%%%%%%

In calculating matrix elements of the boost operator one encounters unusual features. The presence of $x^i$ turns, in the usual momentum representation, in a derivative respect to $k^i$, that affects in the end the $\delta$-functions from the field (anti)commutators. While more than one such momentum-conservation functions appear, the following construction can be found often:
$\int dk F(k) \delta(k) \partial^2_k \delta(k)
$
that we at present ignore how to handle.
If we ignore the term with second derivatives  acting on the Dirac delta functions, we can complete the rest of the calculation by the usual per-parts integration method that systematically transfers derivatives from  the $\delta$-functions to the other functions being integrated.
To proceed we write the overlap as
\begin{eqnarray}
\la \eta_v^\mathfrak{g}\ar \eta_v^\mathfrak{g}\ra&=& (4\pi\alpha_s)^2\left(
\frac{\zeta^2}{2} \right)^2
\sin^2\left(\phi_\mathrm{P}\right) \int \dbar \mathrm{q} \int \dbar\mathrm{p} \int \dbar \mathrm{q}' \int \dbar \mathrm{p}' \nonumber\\
&\times&\frac{1}{4\pi} \frac{1}{4\mathrm{N}_c} \frac{\epsilon^{ijl}\epsilon^{i'j'l'}}{2\mathrm{E}_\mathbf{p}\mathrm{E}_\mathbf{q}\mathrm{E}_\mathbf{p}'\mathrm{E}_\mathbf{q}'} \mathscr{F}_{\eta_0}^*(\vert\mathbf{q}\vert)\mathscr{F}_{\eta_0}(\vert\mathbf{q}'\vert)\\ \nonumber&\times&
(\mathrm{m}(\vert\mathbf{p}\vert)\mathrm{q}^i-\mathrm{m}(\vert\mathbf{q}\vert)\mathrm{p}^i)  (\mathrm{m}(\vert\mathbf{p}'\vert)\mathrm{q}^{'i'}-\mathrm{m}(\vert\mathbf{q}'\vert)\mathrm{p}^{'i'})\\ \nonumber&\times&
\int \frac{\mathrm{d}^3\mathrm{k}_1\mathrm{d}^3\mathrm{k}_2\mathrm{d}^3\mathrm{k}_1'\mathrm{d}^3\mathrm{k}_2'
}{\left(2\omega_1\omega_2\omega_ 1'\omega_2'\right)^{1/2}}
\partial_{\mathrm{k}_1^z} \delta^{(3)}(\mathbf{k}_1+\mathbf{q}-\mathbf{p})\\ \nonumber&\times&
\partial_{\mathrm{k}_2^z} \delta^{(3)}(\mathbf{k_2}+\mathbf{p}-\mathbf{q}) 
\partial_{\mathrm{k}_1^{'z}} \delta^{(3)}(\mathbf{k'_1}+\mathbf{q'}-\mathbf{p'})\\ \nonumber&\times&
\partial_{\mathrm{k}_2^{'z}} \delta^{(3)}(\mathbf{k'_2}+\mathbf{p'}-\mathbf{q'}) (2\pi)^6\left\{
\delta^{(3)}(\mathbf{k_1}-\mathbf{k_1'}) \right.\\ \nonumber&\times&\textbf{t}^{jj'}(\textbf{k}_1)
\delta^{(3)}(\mathbf{k_2}-\mathbf{k_2'}) \textbf{t}^{ll'}(\textbf{k}_2) + 
\delta^{(3)}(\mathbf{k_1}-\mathbf{k_2'})\\ \nonumber&\times&\textbf{t}^{jl'}(\textbf{k}_1) 
\delta^{(3)}(\mathbf{k_2}-\mathbf{k_1'})\left.\textbf{t}^{lj'}(\textbf{k}_2) \right\}.
\end{eqnarray}
There are several things to remark concerning this expression. First, note that $\textbf{t}^{ij}(\mathbf{k})\equiv (\delta^{ij}-\mathrm{k}^i\mathrm{k}^j/\mathbf{k}^2)$ is the transverse projector in momentum space which ensures transversality. Since the suite of Kronecker deltas will force $\mathbf{k}_1=-\mathbf{k}_2$, and due to the spin contraction with antisymmetric $\epsilon^{ijl}$ symbols,
the terms quartic in $\hat{k}_n$ will not contribute. It is also convenient to define a shorthand notation regrouping the meson wavefunction and the quark propagator pieces,
\begin{equation}
\textbf{G}(\mathbf{p},\mathbf{q}) \equiv 
\mathscr{F}_{\eta_0}(\vert\mathbf{q}\vert) 
\frac{\mathrm{m}(\vert\mathbf{p}\vert)\textbf{q} - \mathrm{m}(\vert\mathbf{q}\vert)\textbf{p}}{\mathrm{E}_\mathbf{p} \mathrm{E}_\mathbf{q}}
\end{equation}
(where the second wavefunction depending on $\vert\mathbf{p}\vert$ arises due to the second contributing spin combination).

Finally, as explained above,
\ba \nonumber
\partial_{\mathrm{k}_1^z} \delta^{(3)}(\mathbf{k}_1+\mathbf{q}-\mathbf{p})
\partial_{\mathrm{k}_2^z} \delta^{(3)}(\mathbf{k}_1+\mathbf{q}-\mathbf{p}) \times [\dots]
\\ \nonumber
\to  \delta^{(3)}(\mathbf{k}_1+\mathbf{q}-\mathbf{p})  \delta^{(3)}(\mathbf{k}_1+\mathbf{q}-\mathbf{p})  \partial_{\mathrm{p}^z} \partial_{\mathrm{q}^z} [\dots]
\ea
ignoring the effect of the derivatives over other delta functions, that would normally lead to the $\delta \nabla^2 \delta$ term. Due to invariance under rotations, one can also substitute derivatives  by the Laplacian $\partial^2/\partial_{k_z}^2 \to \nabla^2/3$.

After isolating the $\delta$'s, one encounters two factors of the type
$$
\frac{1}{9} \nabla^2_{\mathbf{k}_1} \nabla^2_{\mathbf{k}_2} \left\{
\frac{1}{2\omega_{\mathbf{k}_1}2\omega_{\mathbf{k}_2}} \textbf{t}^{jl'}(\textbf{k}_1)
\textbf{t}^{j'l}(\textbf{k}_2)
\right\}$$
that are handled by means of
\begin{eqnarray}
 \nabla^2_{\mathbf{k}} \left\{ \frac{1}{2\omega_{\mathbf{k}}} \textbf{t}^{ij}(\textbf{k})\right\} &=&
\frac{\textbf{t}^{ij}(\textbf{k})}{\omega_\mathbf{k}^2}  \left(
\frac{\left(\omega_\mathbf{k}^{\prime}\right)^2}{4\omega_\mathbf{k}} -\frac{\omega_\mathbf{k}^{''}}{2}-\frac{\omega'}{\ar\mathbf{k} \ar} \right)
\\&-&\frac{2}{\omega\ar\mathbf{k}\ar^2} \left(\delta^{ij}-3\frac{\mathrm{k}^i \mathrm{k}^j}{\mathbf{k}^2}\right) \nonumber .
\end{eqnarray}
Since there are two insertions of the boost operator, this is a three-loop calculation. The integration variables can be chosen as $\mathbf{q}$, $\mathbf{p}$
and $\mathbf{q'}$. The fourth, $\mathbf{p}'$ becomes a linear combination. There are two terms with a different value of $\mathbf{p}'$ in each, that we will denote with a subindex
\begin{eqnarray}\nonumber
\mathbf{p}'_1&=&  \mathbf{p}-\mathbf{q}+\mathbf{q}'\\ \nonumber
\mathbf{p}'_2&=& -\mathbf{p}+\mathbf{q}+\mathbf{q}'\ .
\end{eqnarray}

We can then write down our partial calculation for the overlap as
\ba \label{overlapfinal}
\frac{\la \eta^\mathfrak{g}_v \ar \eta^\mathfrak{g}_v \ra }{\la 0 \ar 0 \ra}&=&
 (4\pi\alpha_s)^2\left(\frac{\zeta^2}{2} \right)^2 \sin^2\left(\phi_\mathrm{P}\right) \frac{1}{4\pi} \frac{1}{4\mathrm{N}_c} \\ \nonumber&\times&
 \int \dbar \mathrm{q}\int \dbar  \mathrm{p}\int \dbar \mathrm{q}'
\frac{1}{9}\left( \mathscr{S}-\tilde{\mathscr{S}} \right)
\ea
with spin work $\mathscr{S}$ that can be obtained by ($\boldsymbol{\Delta}_{\mathbf{q}-\mathbf{p}}\equiv (\textbf{q-p})/\vert\textbf{q-p}\vert$)
\begin{eqnarray}
\frac{\epsilon^{ijl}\epsilon^{i'j'l'}}{2} (\delta^{jj'}-3\Delta^j \Delta^{j'}
) (\delta^{ll'}-3\Delta^l \Delta^{l'})  G^i G^{i'} = \\ \nonumber
-2 \mathbf{G} \cdot \mathbf{G}' + 3 (\mathbf{\Delta} \cdot \mathbf{G}) (\mathbf{\Delta} \cdot\mathbf{G}')
\end{eqnarray}
and similar relations, to yield ($\omega\equiv\omega_{\mathbf{q}-\mathbf{p}}$)
\begin{eqnarray} 
\mathscr{S}&=&\left\{
\left[\frac{1}{\omega^2} \left( \frac{\omega^{'2}}{4\omega}-\frac{\omega^{''}}{2}-\frac{\omega'}{\ar \mathbf{q}-\mathbf{p}\ar}\right)\right]^2\left[\mathbf{G}(\mathbf{p},\mathbf{q})\!\cdot\!\boldsymbol{\Delta}_{\mathbf{q}-\mathbf{p}}\right]\right. \\ \nonumber  &\times&\ \  \left[\mathbf{G}(\mathbf{p}_1',\mathbf{q}')\!\cdot\!\boldsymbol{\Delta}_{\mathbf{q}-\mathbf{p}}\right]
-\frac{2}{\omega \ar \mathbf{q}-\mathbf{p}\ar^2}
\left[\frac{1}{\omega^2} \left( \frac{\omega^{'2}}{4\omega}-\frac{\omega^{''}}{2}\right.\right.\\\nonumber&-&\left.\left.\frac{\omega'}{\ar \mathbf{q}-\mathbf{p}\ar}\right)\right]\left(-\mathbf{G}(\mathbf{p},\mathbf{q})\!\cdot \!\mathbf{G}(\mathbf{p}_1',\mathbf{q}')
+ 2 \left[\mathbf{G}(\mathbf{p},\mathbf{q})\!\cdot\!\boldsymbol{\Delta}_{\mathbf{q}-\mathbf{p}}\right]\right.\\\nonumber&\times&\left.\left[\mathbf{G}(\mathbf{p}_1',\mathbf{q}')\!\cdot \!\boldsymbol{\Delta}_{\mathbf{q}-\mathbf{p}}\right]
\right)+ \frac{4}{\omega^2 \ar \mathbf{q}-\mathbf{p}\ar^4} \left(-2 \mathbf{G}(\mathbf{p},\mathbf{q})\!\cdot\right.\\\nonumber &&\left.\mathbf{G}(\mathbf{p}_1',\mathbf{q}')
+\left.3(\mathbf{G}(\mathbf{p},\mathbf{q})\!\cdot\!\boldsymbol{\Delta}_{\mathbf{q}-\mathbf{p}})(\mathbf{G}(\mathbf{p}_1',\mathbf{q}')\!\cdot \!\boldsymbol{\Delta}_{\mathbf{q}-\mathbf{p}}\right)
\right\}
\end{eqnarray}and with $\tilde{\mathscr{S}}$ obtainable replacing the argument of $\textbf{G}(\mathbf{p}'_1,\mathbf{q}')$ by the second solution to the momentum $\delta$'s  $\textbf{G}(\mathbf{p}_ 2',\mathbf{q}')$.
With all these factors, one can proceed to a numerical evaluation of Eq.~(\ref{overlapfinal}). Given the now explicit invariance under rotations of the integrand, the nine-dimensional integral can be reduced by choosing the $z$ axis along $\bf p$, and the $x$ axis so that $\textbf{p}$, $\textbf{q}$ are in the $xz$ plane. The remaining integration variables are then the three moduli $\vert\textbf{p}\vert$,  $\vert\textbf{q}\vert$, $\vert\textbf{q}'\vert$, and the three integration angles
$\theta_\mathbf{q}$, $\Omega_{\mathbf q'}$.
This six-dimensional integral can be carried out with the help of VEGAS, the standard adaptive Montecarlo algorithm \cite{Lepage}.
The running quark masses $\rm m(\vert\mathbf{q}\vert)$ and gluon (dispersive) energies $\omega(\vert\bf k\vert)$ appearing satisfy their respective gap equations. We take them to be solutions of the truncated versions in
\cite{llanescotanch} and \cite{Szczepaniak:1995cw}, as later work, especially in the gluon sector, has improved the analysis but not changed the numerical results qualitatively.

Exploration of the six-dimensional numerical integral $\Im$, defined by
\begin{equation}
\langle \eta_v^\mathfrak{g} \vert\eta_v^\mathfrak{g}\rangle = (4\pi\alpha_s)^2\left(
\frac{\zeta^2}{2} \right)^2 \sin^2\left(\phi_\mathrm{P}\right) \Im
 \end{equation}
and taking into account conceivable uncertainties in the quarkonium $\eta$ wavefunction, quark mass gap function, and gluon dispersive function, allows us to make an order of magnitude estimate (note $\Im$, after extraction of the volume factor $(2\pi)^3\delta^{(3)}(0)$ is a pure, dimensionless number)
$\log \Im= 2.6^{+0.5}_{-2.0}$.
The prefactor shows the explicit dependence in the pseudoscalar mixing angle (through which the singlet component of the $\eta$ arises), which we take at
39 degrees, the rapidity $\zeta\simeq 0.62$ of the boosted $\eta$ meson, and $\alpha_s$ at some low scale. This last value is not known with certainty, but since it is estimated to be $0.4$ in $\tau$ decays, we obtain a lower bound to the prefactor of $\Im$ which is 0.4.

Putting all together, we obtain
\begin{equation}
\log \langle \eta_v^\mathfrak{g}\vert \eta_v^\mathfrak{g}\rangle = 1.8^{+0.5}_{-2.0} 
\end{equation}
which means that $\langle\eta_v^\mathfrak{g}\vert \eta_v^\mathfrak{g}\rangle$ is larger than $1$ by somewhat more than one sigma. This is of course not possible, as the sum of all coefficients of the wavefunction in Eq.(\ref{Fockexp}) $\sum_i \rm c^2_i=1$, but one has to take into account that this  is only a partial calculation including only one of the relevant terms of the boost operator, and that model uncertainties creep in the choice of wave and gap functions.

What one should conclude is that there is no reason why $\langle \eta_v^\mathfrak{g}\vert \eta_v^\mathfrak{g}\rangle$
should be orders of magnitude smaller than one. By the size of the controled factors one could guess it to be of order 0.1. That is, we believe we can argue with certain confidence that a gluonium component of the $\eta$ meson is induced by boosting it to a moving reference frame, as in $\phi$ radiative decays, even if this component was not present in the rest-frame wavefunction.

%%%%%%%%%%%%%%%%%%%%%%%%%%%%%%%%%%%%%%%%%%%%
\section{Summary}\label{sec:conclusions}
%%%%%%%%%%%%%%%%%%%%%%%%%%%%%%%%%%%%%%%%%%%

In this work we have derived the boost generators of Quantum Chromodynamics quantized in Coulomb gauge in the path integral formalism.
These confirm the result of Besting and Sch\"utte for pure Yang-Mills theory, to which we add the fermion fields to complete the boost operator of canonically quantized QCD in Coulomb gauge
\begin{eqnarray}\label{Boostfinal}
\hat{\mathbf{K}}_{\mathrm{QCD}} &=& 
-\int \mathrm{d}^3 \mathrm{x}\left\{\frac{1}{2}\mathcal{J}^{-1} \hat{\mathbf{\Pi}}^a \mathcal{J} 
\mathbf{x}\hat{\mathbf{\Pi}}^a+\frac{1}{2}\hat{\mathbf{B}}^a \mathbf{x} \hat{\mathbf{B}}^a \right.
\\
&-&\frac{1}{2}\mathrm{g}^2\mathcal{J}^{-1}\hat{\varrho}^{b}\mathcal{J} \hat{\mathrm{M}}^{ba}(\partial_\ell \mathbf{x}\partial_\ell) \hat{\mathrm{M}}^{ac}\hat{\varrho}^c\nonumber
\\ 
&+&\left. \frac{1}{2}\mathrm{g}\mathcal{J}^{-1}\hat{\mathbf{\Pi}}^a \mathcal{J}\hat{\mathrm{M}}^{ab}\hat{\varrho}^b+\frac{1}{2}\mathrm{g} \mathcal{J}^{-1}\hat{\varrho}^{b}\mathcal{J} \hat{\mathrm{M}}^{ba}\hat{\mathbf{\Pi}}^a\right\}
\nonumber \\
&-&\int \mathrm{d}^3\mathrm{x} \mathbf{x} \left\{\hat{\mathfrak{q}}^{\ell\dagger} (\mathbf{x}) \left( -i\mathbf{\alpha}\cdot \mathbf{\nabla} + \beta\mathrm{m}_\mathfrak{q} \right) \hat{\mathfrak{q}}^\ell(\mathbf{x})\right.\nonumber\\ 
&+&\left. \mathrm{g}\hat{\mathfrak{q}}^{\ell\dagger} (\mathbf{x}) \frac{\lambda^a}{2}\mathbf{\alpha}\cdot \hat{\mathbf{A}}^a(\mathbf{x}) \hat{\mathfrak{q}}^\ell(\mathbf{x})\right\}\nonumber
\\
&+&\int \mathrm{d}^3 \mathrm{x}\hat{\mathfrak{q}}^{\ell\dagger}(\mathbf{x})\left(\frac{i}{2}\mathbf{\alpha} \right) \hat{\mathfrak{q}}^\ell (\mathbf{x}) \ . \nonumber
\end{eqnarray}

We have wanted to call the attention of the community to the dynamical nature of these operators. We have focused on two aspects of current interest for the interpretation of hadron data in terms of their quark and gluon constituents.
The first is the lack of a charge-distribution interpretation of the pion form factor. The operator matrix elements that one obtains from such experiments are a much more complicated generalization.
The second is the fact that Fock-space expansions of hadron wavefunctions are tied to the rest frame of the hadron, and they change with its velocity.  This is of current interest for the theoretical understanding of pseudoscalar mesons produced in $\phi$ radiative decays at Frascatti.

Many more applications come to mind where theoretical interpretation is complicated by the dynamical nature of the boost operators. However there is no known practical way of handling them to obtain even model results. We look forward to any progress in this direction by lattice methods or other means.
\footnote{Of course, Light Front quantization or explicitly covariant approaches such as the Dyson-Schwinger equations are free of this problem by construction but they have other issues of their own that need to be overcome. Coulomb gauge QCD remains an attractive avenue of thought for non-perturbative problems since the physical Hilbert space makes transparent the application of the variational principle.}
 \vspace{0.20 in}
\begin{flushleft}
\textbf{\large{Acknowledgments}}
\end{flushleft}
\vspace{0.10 in}
\emph{Work supported by grants FPA 2008-00592, FIS2008-01323,
FPA2007-29115-E, PR34-1856-BSCH, UCM-BSCH GR58/08 910309, PR34/07-15875 (Spain).}  S.~Villalba-Ch\'avez is  supported by the Doktoratskolleg ``Hadrons in Vacuum, Nuclei and Stars''  of the Austrian science fund (FWF) under contract W1203-N08. This author  and Dieter Sch\"utte thank the hospitality of the group at Universidad Complutense where this work was completed. Part of it has been presented to the University in the Master's thesis of Maria G\'omez Rocha.
%%%%%%%%%%%%%%%%%%%%%%%%%%%%%%%%%%%%%%%%%%%%

%%%%%%%%%%%%%%%%%%%%%%%%%%%%%%%%%%%%%%%%%%%%%%%%%%%%%%%%%%
\appendix
%%%%%%%%%%%%%%%%%%%%%%%%%%%%%%%%%%%%%%%%%%%%%%%%%%%%%%%%%%
\section{The Fadeev-Popov Determinant and the Hermiticity of $\hat{\mathbf{\Pi}}^a(\mathbf{x}).$\label{fadeevandhermeticity}}
Let us first recheck for the reader not familiar with functional quantization how Hermiticity of the canonical momentum comes about, in total analogy with elementary quantum mechanics,
\begin{equation}
\langle\Psi_1\vert\hat{\mathbf{\Pi}}^a(\mathbf{x})\vert\Psi_2\rangle=\int\mathcal{D}\mathbf{A}\mathcal{J}\Psi_1^*[\mathbf{A}]\frac{1}{i}\frac{\delta}{\delta\hat{\mathbf{A}}^a(\mathbf{x})}\Psi_2[\mathbf{A}]\ .
\end{equation}
Again integrating by parts
\begin{eqnarray}
\langle\Psi_1\vert\hat{\mathbf{\Pi}}^a(\mathbf{x})\vert\Psi_2\rangle&=&-\int\mathcal{D}\mathbf{A}\left[\frac{1}{i}\frac{\delta}{\delta\hat{\mathbf{A}}^a(\mathbf{x})}\mathcal{J}\Psi_1^*[\mathbf{A}]\right]\Psi_2[\mathbf{A}]\nonumber\\&=&
\int\mathcal{D}\mathbf{A}\left[\hat{\mathbf{\Pi}}^{a}(\mathbf{x})\mathcal{J}\Psi_1[\mathbf{A}]\right]^{*}\Psi_2[\mathbf{A}]\label{trucohermitcvgappend}\ .
\end{eqnarray}
The introduction of a unit factor $1=\mathcal{J}\mathcal{J}^{-1}$  in front of the bracket of the above expression  leads to 
\begin{eqnarray}\label{trucohermitcvgappend2}
\langle\Psi_1\vert\hat{\mathbf{\Pi}}^a(\mathbf{x})\vert\Psi_2\rangle=
\int\mathcal{D}\mathbf{A}\mathcal{J}\left[\mathcal{J}^{-1}\hat{\mathbf{\Pi}}^{a}(\mathbf{x})\mathcal{J}\Psi_1[\mathbf{A}]\right]^{*}\Psi_2[\mathbf{A}]\ .\nonumber\\
\end{eqnarray}
 It is to notice that  
\begin{eqnarray}\label{defhermconasdr}
\langle\Psi_1\vert\hat{\mathbf{\Pi}}^{a\dagger}(\mathbf{x})\vert\Psi_2\rangle&=&\langle\hat{\mathbf{\Pi}}^{a}(\mathbf{x})\Psi_1\vert\Psi_2\rangle\\&=&\int\mathcal{D}\mathbf{A}\mathcal{J}\left[\hat{\mathbf{\Pi}}^{a\dagger}(\mathbf{x})\Psi_1[\mathbf{A}]\right]^{*}\Psi_2[\mathbf{A}]\nonumber
\end{eqnarray} 
and we compare  Eq. (\ref{trucohermitcvgappend2}) with Eq. (\ref{defhermconasdr}) to  find that with the curvilinear scalar product, the canonical momentum is not Hermitian, $
\hat{\mathbf{\Pi}}^{a\dagger}(\mathbf{x})=\mathcal{J}^{-1}\hat{\mathbf{\Pi}}^{a}(\mathbf{x})\mathcal{J}.$

%%%%%%%%%%%%%%%%%%%%%%%%%%%%%%%%%%%%%%%%%%%%%%%%%%%%%%%%%%%%%%%%%%%%%%%%%%%%


\begin{thebibliography}{99}
%%%%%%%%%%%%%%%%%%%%%%%%%%%%%%%%%%%%%%%%%%%%%%%%%%%%%%%%%%%%%%%%%%%%%%%%%%%%

\bibitem{hofstaedter}
 M.~E.~Rose,
  %``The Charge Distribution in Nuclei and the Scattering of High Energy
  %Electrons,''
  Phys.\ Rev.\  {\bf 73} (1948) 279.

%\cite{Gasser:1990bv}
\bibitem{Gasser:1990bv}
  J.~Gasser and U.-G.~Mei\ss ner,
  %``Chiral expansion of pion form-factors beyond one loop,''
  Nucl.\ Phys.\  B {\bf 357} (1991) 90.
  %%CITATION = NUPHA,B357,90;%%
%\cite{Guo:2008nc}
\bibitem{Guo:2008nc}
  F.~K.~Guo, C.~Hanhart, F.~J.~Llanes-Estrada and U.~G.~Meissner,
  %``Quark mass dependence of the pion form factor,''
  Phys.\ Lett.\  B {\bf 678} (2009) 90
  [arXiv:0812.3270 [hep-ph]].
  %%CITATION = PHLTA,B678,90;%%

%\cite{Amendolia:1986wj}
\bibitem{Amendolia:1986wj}
  S.~R.~Amendolia {\it et al.}  [NA7 Collaboration],
  %``A Measurement Of The Space - Like Pion Electromagnetic Form-Factor,''
  Nucl.\ Phys.\  B {\bf 277} (1986) 168.
  %%CITATION = NUPHA,B277,168;%%


%\cite{Huber:1998zz}
\bibitem{Huber:1998zz}
  A.~Huber {\it et al.},
  %``Hydrogen-Deuterium S-1- S-2 Isotope Shift and the Structure of the
  %Deuteron,''
  Phys.\ Rev.\ Lett.\  {\bf 80}, 468 (1998).
  %%CITATION = PRLTA,80,468;%%


%\cite{Friar:1997js}
\bibitem{Friar:1997js}
  J.~L.~Friar, J.~Martorell and D.~W.~L.~Sprung,
  %``Nuclear Sizes and the Isotope Shift,''
  Phys.\ Rev.\  A {\bf 56}, 4579 (1997)
  [arXiv:nucl-th/9707016].
  %%CITATION = PHRVA,A56,4579;%%

\bibitem{Friar:1979ak}
  J.~L.~Friar,
  %``Corrections To The Impulse Approximation. 1. Dispersion And Recoil Effects
  %In Elastic Electron Scattering. 2. Mesboldsymbol{\mathscron Exchange Currents And Relativistic
  %Effects,''
  Ann. Phys. {\bf 81} (1973) 332-363.
  %%CITATION = C79-06-18.2-3;%%

\bibitem{Brodsky:1968ea}
  S.~J.~Brodsky and J.~R.~Primack,
  %``The Electromagnetic Interactions Of Composite Systems,''
  Annals Phys.\  {\bf 52}, 315 (1969).
  %%CITATION = APNYA,52,315;%%

%\cite{Choi:2008yj}
\bibitem{Choi:2008yj}
  H.~M.~Choi and C.~R.~Ji,
  %``Conformal Symmetry and Pion Form Factor: Space- and Time-like Region,''
  Phys.\ Rev.\  D {\bf 77} (2008) 113004
  [arXiv:0803.2604 [hep-ph]].
  %%CITATION = PHRVA,D77,113004;%%



\bibitem{Biernat:2009my}
  E.~P.~Biernat, W.~Schweiger, K.~Fuchsberger and W.~H.~Klink,
  %``Electromagnetic meson form factor from a relativistic coupled-channel
  %approach,''
  arXiv:0902.2348 [nucl-th].
  %%CITATION = ARXIV:0902.2348;%%

%\cite{Plessas:2002vq}
\bibitem{Plessas:2002vq}
  W.~Plessas, S.~Boffi, L.~Y.~Glozman, W.~Klink, M.~Radici and R.~F.~Wagenbrunn,
  %``Nucleon properties in a semirelativistic chiral quark model,''
  Nucl.\ Phys.\  A {\bf 699}, 312 (2002).
  %%CITATION = NUPHA,A699,312;%%


\bibitem{Brodsky:2007hb}
  S.~J.~Brodsky and G.~F.~de Teramond,
  %``Light-Front Dynamics and AdS/QCD Correspondence: The Pion Form Factor in
  %the Space- and Time-Like Regions,''
  Phys.\ Rev.\  D {\bf 77} (2008) 056007
  [arXiv:0707.3859 [hep-ph]].
  %%CITATION = PHRVA,D77,056007;%%

%\cite{Szczepaniak:1995mi}
\bibitem{Szczepaniak:1995mi}
  A.~Szczepaniak, C.~R.~Ji and S.~R.~Cotanch,
  %``Nucleon structure in a relativistic quark model,''
  Phys.\ Rev.\  C {\bf 52} (1995) 2738.
  %%CITATION = PHRVA,C52,2738;%%

%\cite{DiMicco:2009zza}
\bibitem{DiMicco:2009zza}
  B.~Di Micco  [KLOE Collaboration],
  %``Measurement Of The Pseudoscalar Mixing Angle And The Eta-Prime Gluonium
  %Content With The Kloe Detector,''
  Acta Phys.\ Polon.\ Supp.\  {\bf 2} (2009) 63;
  %%CITATION = APPXA,2,63;%%
  B.~Di Micco  [KLOE Collaboration],
  %``eta, eta-prime mixing angle and eta-prime gluonium content extraction from
  %the KLOE R(phi) measurement,''
  Eur.\ Phys.\ J.\  A {\bf 38} (2008) 129.
  %%CITATION = EPHJA,A38,129;%%

%\cite{Bramon:2000fr}
\bibitem{Bramon:2000fr}
  A.~Bramon, R.~Escribano and M.~D.~Scadron,
  %``Radiative V P gamma transitions and eta eta' mixing,''
  Phys.\ Lett.\  B {\bf 503} (2001) 271
  [arXiv:hep-ph/0012049].
  %%CITATION = PHLTA,B503,271;%%



%\cite{Escribano:2007cd}
\bibitem{Escribano:2007cd}
  R.~Escribano and J.~Nadal,
  %``On the gluon content of the eta and eta' mesons,''
  JHEP {\bf 0705} (2007) 006
  [arXiv:hep-ph/0703187].
  %%CITATION = JHEPA,0705,006;%%


%\cite{Escribano:2008rq}
\bibitem{Escribano:2008rq}
  R.~Escribano,
  %``J/psi->VP decays and the quark and gluon content of the eta and eta',''
  arXiv:0807.4201 [hep-ph];
  %%CITATION = ARXIV:0807.4201;%%
  R.~Escribano,
  %``Eta' gluonic content and J/psi->VP decays,''
  Nucl.\ Phys.\ Proc.\ Suppl.\  {\bf 181-182} (2008) 226
  [arXiv:0807.4205 [hep-ph]].
  %%CITATION = NUPHZ,181-182,226;%%


%\cite{Maris:2000sk}
\bibitem{Maris:2000sk}
  P.~Maris and P.~C.~Tandy,
  %``The pi, K+, and K0 electromagnetic form factors,''
  Phys.\ Rev.\  C {\bf 62} (2000) 055204
  [arXiv:nucl-th/0005015].
  %%CITATION = PHRVA,C62,055204;%%


%\cite{Sachs:1962zzc}
\bibitem{Sachs:1962zzc}
  R.~G.~Sachs,
  %``High-Energy Behavior of Nucleon Electromagnetic Form Factors,''
  Phys.\ Rev.\  {\bf 126} (1962) 2256.
  %%CITATION = PHRVA,126,2256;%%
%\cite{Kelly:2002if}
\bibitem{Kelly:2002if}
  J.~J.~Kelly,
  %``Nucleon charge and magnetization densities from Sachs form factors,''
  Phys.\ Rev.\  C {\bf 66}, 065203 (2002)
  [arXiv:hep-ph/0204239].
  %%CITATION = PHRVA,C66,065203;%%

%\cite{Amsler:2008zzb}
\bibitem{Amsler:2008zzb}
  C.~Amsler {\it et al.}  [Particle Data Group],
  %``Review of particle physics,''
  Phys.\ Lett.\  B {\bf 667} (2008) 1.
  %%CITATION = PHLTA,B667,1;%%


\bibitem{latticepapers}
  P.~A.~Boyle {\it et al.},
  %``The pion's electromagnetic form factor at small momentum transfer in full
  %lattice QCD,''
  JHEP {\bf 0807} (2008) 112
  [arXiv:0804.3971 [hep-lat]];
  %%CITATION = JHEPA,0807,112;%%
  S. Aoki {\it et al.}  [TWQCD collaboration],
  %``Pion form factors from two-flavor lattice QCD with exact chiral symmetry,''
  arXiv:0905.2465 [hep-lat];
  %%CITATION = ARXIV:0905.2465;%%
  R.~Frezzotti, V.~Lubicz and S.~Simula,
  %``Electromagnetic form factor of the pion from twisted-mass lattice QCD at
  %Nf=2,''
  arXiv:0812.4042 [hep-lat].
  %%CITATION = ARXIV:0812.4042;%%


%\cite{VillalbaChavez:2008dv}
\bibitem{VillalbaChavez:2008dv}
  S.~Villalba-Chavez, R.~Alkofer and K.~Schwenzer,
  %``On the connection between Hamilton and Lagrange formalism in Quantum Field
  %Theory,''
  arXiv:0807.2146 [hep-th].
  %%CITATION = ARXIV:0807.2146;%%

%\cite{Zwanziger:1998ez}
\bibitem{Zwanziger:1998ez}
  D.~Zwanziger,
  %``Renormalization in the Coulomb gauge and order parameter for  confinement
  %in QCD,''
  Nucl.\ Phys.\  B {\bf 518} (1998) 237.
  %%CITATION = NUPHA,B518,237;%%

%\cite{Watson:2006yq}
\bibitem{Watson:2006yq}
  P.~Watson and H.~Reinhardt,
  %``Propagator Dyson-Schwinger equations of Coulomb gauge Yang-Mills theory
  %within the first order formalism,''
  Phys.\ Rev.\  D {\bf 75} (2007) 045021
  [arXiv:hep-th/0612114].
  %%CITATION = PHRVA,D75,045021;%%


%\cite{Besting:1989nq}
\bibitem{Besting:1989nq}
  P.~Besting and D.~Schutte,
  %``RELATIVISTIC INVARIANCE OF COULOMB GAUGE YANG-MILLS THEORY,''
  Phys.\ Rev.\  D {\bf 42} (1990) 594.
  %%CITATION = PHRVA,D42,594;%%

%\GribovWM
\bibitem{gribov}
  V.~N.~Gribov,
  %``Quantization of non-Abelian gauge theories,''
  Nucl.\ Phys.\  B {\bf 139}, 1 (1978).
  %%CITATION = NUPHA,B139,1;%%

%\ZwanzigerCF
\bibitem{zwanziger2}
  D.~Zwanziger,
  %``Non-perturbative Faddeev-Popov formula and infrared limit of QCD,''
  Phys.\ Rev.\  D {\bf 69}, 016002 (2004)
  [arXiv:hep-ph/0303028].
  %%CITATION = PHRVA,D69,016002;%%


%\cite{Reinhardt:2008pr}
\bibitem{Reinhardt:2008pr}
  H.~Reinhardt and P.~Watson,
  %``Resolving temporal Gribov copies in Coulomb gauge Yang-Mills theory,''
  Phys.\ Rev.\  D {\bf 79} (2009) 045013
  [arXiv:0808.2436 [hep-th]].
  %%CITATION = PHRVA,D79,045013;%%


%\cite{Christ:1980ku}
\bibitem{Christ:1980ku}
  N.~H.~Christ and T.~D.~Lee,
  %``Operator Ordering And Feynman Rules In Gauge Theories,''
  Phys.\ Rev.\  D {\bf 22} (1980) 939
  [Phys.\ Scripta {\bf 23} (1981) 970].
  %%CITATION = PHSTB,23,970;%%
%\cite{Bicudo:2006sd}

%\cite{Feuchter:2004mk}
\bibitem{Feuchter:2004mk}
  C.~Feuchter and H.~Reinhardt,
  %``Variational solution of the Yang-Mills Schroedinger equation in Coulomb
  %gauge,''
  Phys.\ Rev.\  D {\bf 70} (2004) 105021
  [arXiv:hep-th/0408236];
  %%CITATION = PHRVA,D70,105021;%%
 
%\cite{Szczepaniak:2001rg}
\bibitem{Szczepaniak:2001rg}
  A.~P.~Szczepaniak and E.~S.~Swanson,
  %``Coulomb gauge QCD, confinement, and the constituent representation,''
  Phys.\ Rev.\  D {\bf 65} (2002) 025012
  [arXiv:hep-ph/0107078].
  %%CITATION = PHRVA,D65,025012;%%

\bibitem{Szczepaniak:2003ve}
 A.~P.~Szczepaniak,
  %``Confinement and gluon propagator in Coulomb gauge QCD,''
  Phys.\ Rev.\  D {\bf 69} (2004) 074031
  [arXiv:hep-ph/0306030].


%\cite{Campagnari:2009km}
\bibitem{Campagnari:2009km}
  D.~R.~Campagnari, H.~Reinhardt and A.~Weber,
  %``Perturbation theory in the Hamiltonian approach to Yang-Mills theory in
  %Coulomb gauge,''
  Phys.\ Rev.\  D {\bf 80} (2009) 025005
  [arXiv:0904.3490 [hep-th]].
  %%CITATION = PHRVA,D80,025005;%%
  
%\cite{Watson:2008fb}
\bibitem{Watson:2008fb}
  P.~Watson and H.~Reinhardt,
  %``Slavnov-Taylor identities in Coulomb gauge Yang-Mills theory,''
  arXiv:0812.1989 [hep-th].
  %%CITATION = ARXIV:0812.1989;%%

\bibitem{Bicudo:2006sd}
  P.~Bicudo, S.~R.~Cotanch, F.~J.~Llanes-Estrada and D.~G.~Robertson,
  %``The new f0(1810) in omega Phi at BES: A good glueball candidate,''
  Eur.\ Phys.\ J.\  C {\bf 52} (2007) 363
  [arXiv:hep-ph/0602172].
  %%CITATION = EPHJA,C52,363;%%


\bibitem{Lepage}
  G.~P.~Lepage, CLNS-80/447, 1980.
  %``Vegas: An Adaptive Multidimensional Integration Program,''
  %%CITATION = CLNS-80/447;%%

\bibitem{llanescotanch}
  F.~J.~Llanes-Estrada and S.~R.~Cotanch,
  %``Relativistic many-body Hamiltonian approach to mesons,''
  Nucl.\ Phys.\  A {\bf 697} (2002) 303
  [arXiv:hep-ph/0101078].
  %%CITATION = NUPHA,A697,303;%%


%\cite{Szczepaniak:1995cw}
\bibitem{Szczepaniak:1995cw}
  A.~Szczepaniak, E.~S.~Swanson, C.~R.~Ji and S.~R.~Cotanch,
  %``Glueball Spectroscopy in a Relativistic Many-Body Approach to Hadron
  %Structure,''
  Phys.\ Rev.\ Lett.\  {\bf 76} (1996) 2011
  [arXiv:hep-ph/9511422].
  %%CITATION = PRLTA,76,2011;%%



\end{thebibliography}
\end{document}